\def\Iml{\Im (\Lambda)}
\def\slashchar#1{\setbox0=\hbox{$#1$}           
   \dimen0=\wd0                                 
   \setbox1=\hbox{/} \dimen1=\wd1               
   \ifdim\dimen0>\dimen1                        
      \rlap{\hbox to \dimen0{\hfil/\hfil}}      
      #1                                        
   \else                                        
      \rlap{\hbox to \dimen1{\hfil$#1$\hfil}}   
      /                                         
   \fi}                                         %
\long\def\one#1#2#3
{
\begin{figure}
   \centering \leavevmode
   \epsfxsize=3.5in
   \epsfbox{#1}
   \vskip 0.3in
   \caption{\label{fig:#2} \normalsize #3}
   \hfill
\end{figure}
}

\long\def\ones#1#2#3
{
\begin{figure}
   \centering \leavevmode
   \epsfxsize=3.4in
   \epsfbox{#1}
   \vskip 0.3in
   \caption{\label{fig:#2} \normalsize #3}
   \hfill
\end{figure}
}
\long\def\prdfig#1#2#3
{
\begin{figure}
   \centering \leavevmode
   \epsfxsize=5.5in
   \epsfbox{#1}
   \vskip 0.3in
   \caption{\label{fig:#2} \normalsize #3}
   \hfill
\end{figure}
}

\documentstyle[aps,prl,preprint,floats,epsfig]{revtex}
\begin{document}
\tighten
\preprint{\tighten\vbox{\hbox{\hfil CLNS 01/1730}
                        \hbox{\hfil CLEO 01-6}
}}
\title{Search for CP violation in $\tau\to\pi\pi^0\nu_{\tau}$ decay.}  
\author{(CLEO Collaboration)}
\date{April 3, 2000}
\maketitle

\begin{abstract} 
We search for CP non-conservation in the decays of $\tau$ leptons 
produced via $e^+e^-$ annihilation at $\sqrt{s}\sim 10.6\;$GeV. 
The method uses correlated decays of pairs of $\tau$ 
leptons, each decaying to the $\pi \pi^0 \nu_{\tau}$ final state.  
The search is done within the framework of a model with a
scalar boson exchange.  In an analysis of a data sample 
corresponding to 12.2~million produced $\tau$ pairs  
collected with the CLEO detector, 
we find no evidence of violation of CP symmetry. 
We obtain a limit on the imaginary part of the coupling constant 
parameterizing the relative contribution of diagrams that would lead 
to CP violation to be $-0.046 <\Im(\Lambda) < 0.022$ at $90\%$~C.L. 
This result provides a restriction on CP non-conservation in the 
tau lepton decays.
As a cross check, we study the decay angular distribution and
perform a model-independent search for a CP violation effect of a 
scalar exchange in single $\tau\to \pi\pi^0\nu_\tau$ decays. 
The limit on the imaginary part of the $\tau$ scalar coupling 
is $-0.033 < \Im(\Lambda) < 0.089$ at 90\%~ C.L.
\end{abstract}

\begin{center}
P.~Avery,$^{1}$ C.~Prescott,$^{1}$ A.~I.~Rubiera,$^{1}$
H.~Stoeck,$^{1}$ J.~Yelton,$^{1}$
G.~Brandenburg,$^{2}$ A.~Ershov,$^{2}$ D.~Y.-J.~Kim,$^{2}$
R.~Wilson,$^{2}$
T.~Bergfeld,$^{3}$
% ...
\end{center}
 
\small
\begin{center}
$^{1}${University of Florida, Gainesville, Florida 32611}\\
$^{2}${Harvard University, Cambridge, Massachusetts 02138}\\
$^{3}${University of Illinois, Urbana-Champaign, Illinois 61801}\\
% ...
$^{12}${Purdue University, West Lafayette, Indiana 47907}\\
$^{13}${University of Rochester, Rochester, New York 14627}\\
$^{14}${Southern Methodist University, Dallas, Texas 75275}\\
$^{15}${Syracuse University, Syracuse, New York 13244}\\
% ...
$^{20}${Wayne State University, Detroit, Michigan 48202}\\
% ...
\end{center}
\newpage

\section{Introduction}
\label{intro}

Violation of the combined symmetry of charge conjugation 
and parity (CP) has been of longstanding interest 
as a possible source of the matter-antimatter asymmetry~\cite{matter} 
in the Universe.
Efforts to search for CP-violating effects have concentrated 
so far on the hadronic sector:   
CP violation in strange meson decay has been the subject of intensive  
investigation since its first observation in 1964~\cite{cpk}.
Recent studies of hadronic~\cite{direct} as well as 
semileptonic~\cite{kpimu1,kpimu2,bigi} kaon decays 
provide precision measurements of the CP violation parameters.
Searches for corresponding asymmetries in $B$~meson decays are
the focus of several large ongoing experiments~\cite{BABAR,BELLE}. 
Recent indications of possible neutrino oscillations~\cite{kamiokande} 
make it important to re-examine the question of CP non-conservation 
in the leptonic sector. Such violation is forbidden in the Standard Model 
but appears as a consequence of various extensions~\cite{wudka}.
Models predicting lepton flavor violation often also predict CP
violation in lepton decays~\cite{lfv1,lfv2}.
Among theoretically best-known are the 
multi-Higgs-doublet models (MHDM)~\cite{mhdm,mhdm2,mhdm3}. 
In this paper, we study CP violation in $\tau$ decay, 
in the context of a model with scalar boson exchange~\cite{tsai,kuhn}. 
The results of the search are general but are easiest to 
interpret for a specific choice of MHDM. 
Precision studies of muon decay parameters~\cite{mu_decays_work,mu_decays} show
no indication for a CP violation in such decay. Previous attempts
to study this question in tau decays \cite{colin,belle_result} 
provide only weak restrictions on the CP violation parameters.

The search is carried out using data collected with the CLEO 
detector operating at the Cornell Electron Storage Ring (CESR), 
where $\tau$ leptons are produced in pairs
$e^+e^-\to\tau^+\tau^-$.  
Specifically, we study events in which both $\tau$'s decay
to the $\pi\pi^0\nu_\tau$ final state.  In such events, interference 
between the Standard Model process involving $W$-boson exchange and a 
non-Standard Model one involving scalar boson exchange can give 
rise to observable CP-violating terms.  For each event we compute 
a quantity for which a non-zero expectation value would 
constitute evidence for CP violation.  

This article is organized as follows. 
A general description of CP violation in the leptonic sector is given 
in Section~\ref{cp_violation}. An observable used to search for CP violation
is described in Section~\ref{ss_method}. The data analysis and its results are
in Sections \ref{analysis} and \ref{results}. A model-independent search 
for scalar-mediated $\tau$ decays is described in Section~\ref{helicity}.
Derivations of elements of the probability density distribution for 
a pair of $\tau$ leptons each decaying into $\pi\pi^0\nu_{\tau}$ final states
and theoretical calculations used in optimizing the choice of the 
CP-sensitive variable are given in appendices.

\section{CP violation in lepton decay}
\label{cp_violation}

CP violation generates a difference between the partial decay width for 
a process $(i \to f)$ and the corresponding width for its CP-conjugated 
process $(i^{CP} \to f^{CP})$. Any kinematical observable $\xi$ associated 
with a decay can be described as a sum of CP-even and CP-odd components:
\begin{equation}
\xi = \xi_{even} + \xi_{odd}.
\end{equation}
The average value of this observable is
equal to an integral of $\xi$ over all available phase space 
multiplied by the probability density, $P$, for the decay. 
This probability density also can be, in general, decomposed into
CP-even and CP-odd components:  
\begin{equation}
P = P_{even} + P_{odd}.
\end{equation}
The average value of $\xi$ is: 
$$ <\xi>_{i\to f} = \int (\xi_{even} + \xi_{odd}) P dLips =
           \int (\xi_{even} + \xi_{odd})(P_{even} + P_{odd}) dLips $$
\begin{equation}
 =\int \xi_{even} P_{even} dLips + \int \xi_{odd} P_{odd} dLips.
\end{equation}
Under CP conjugation $P_{odd}$ changes sign and the average value
of the observable $\xi$ for the decay of the CP-conjugated state is:
$$ <\xi>_{i^{CP}\to f^{CP}} = \int (\xi_{even} + \xi_{odd}) P^{CP} dLips =
           \int (\xi_{even} + \xi_{odd})(P_{even} - P_{odd}) dLips $$
\begin{equation}
 =\int \xi_{even} P_{even} dLips - \int \xi_{odd} P_{odd} dLips. 
\end{equation}
If $P_{odd}$ is not equal to zero, then 
$<\xi>_{i\to f} \neq <\xi>_{i^{CP}\to f^{CP}}$ and CP is violated. \\
\vspace*{1cm}
\prdfig{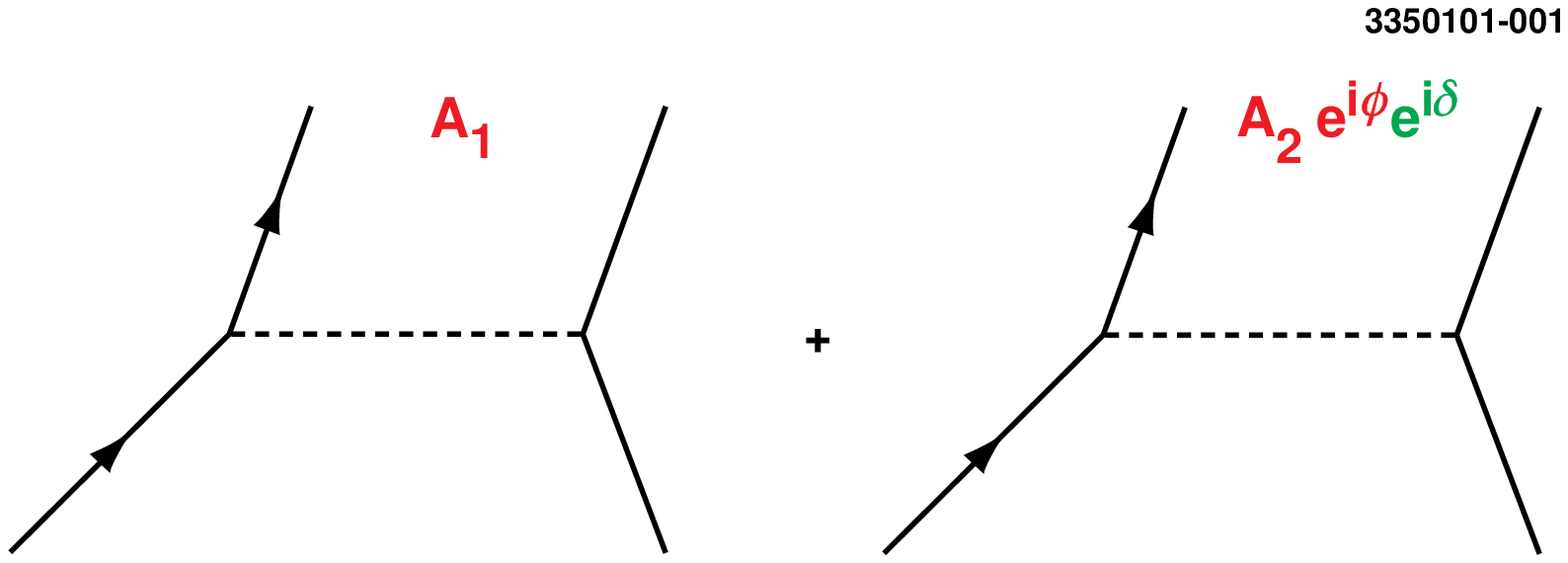}{example}{Interference 
between two amplitudes with CP-even and CP-odd relative phases 
$\delta$ and $\phi$.} 

A lepton decay process described by two amplitudes illustrated in 
Fig.~\ref{fig:example} may have CP-even and CP-odd phases 
$\delta$ and $\phi$ relative to each other. 
The probability density for such a process is given by:
$$ |{\cal A}|^2 = (A_1 + A_2 e^{i\phi} e^{i\delta})
                  (A_1 + A_2 e^{-i\phi} e^{-i\delta})$$
\begin{equation}
 = A_1^2 + A_2^2 + 2 A_1 A_2 \cos \phi \cos \delta - 
          \underline{2 A_1 A_2 \sin \phi \sin \delta}.
\end{equation}
The last, underlined, term is CP-odd since the phase $\phi$ changes
sign under CP conjugation.
In this example the CP-odd term is not equal to zero if
$A_1$, $A_2$, $\sin \phi$, and $\sin \delta$ are not equal to zero.
$A_1$ and $A_2$ denote the amplitudes and for physical processes they
must be different from zero.  Thus CP-odd term is not equal to zero
if the factors $e^{i\phi}$ and $e^{i\delta}$ differ from zero and are complex.
We discuss in the following section a theoretical model that
satisfies these requirements. 

\subsection{\boldmath CP violation in $\tau$ decays}
\label{theory}

A possible scenario for CP violation in $\tau$ lepton decays is
described~\cite{mhdm} by the interference of the Standard Model decay 
amplitude mediated by the W boson (amplitude $A_W$) with the amplitude
mediated by the charged Higgs boson in the multi-Higgs-doublet 
model\footnote{The three-Higgs-doublet model (3HDM) is the least
complicated extension of the Standard Model allowing for CP violation 
in $\tau$ decays~\cite{3hdm1}.} (amplitude $A_H$). These amplitudes play
the roles of $A_1$ and $A_2$ of the previous section.
In this scenario, the charged Higgs couples to quarks and leptons with 
complex coupling constants and, thus, there can be a weak complex 
(CP violating) phase ($\sin \phi \neq 0$). The usual choice
for the CP-even phase $\delta$ is a strong phase~\cite{tsai2} which arises due 
to the QCD final state interactions between quarks. In the following, 
we consider only $\tau$ decays into hadronic final states and a neutrino.
The $\tau$ decay process is described by a sum of two amplitudes: 
Standard Model $W$ exchange illustrated in Fig.~\ref{fig:Aw}(a) and 
a scalar exchange illustrated in Fig.~\ref{fig:Aw}(b).  
\par
To maximize our sensitivity to possible CP-violating effects, 
we optimize our experimental procedures in the context of a specific 
model that allows us to calculate the matrix element and the 
probability density function for $\tau$ decay~\cite{alan}.  We have 
elected to study $\tau$ decay into the $\pi^\pm\pi^0\nu_{\tau}$ final 
state due to its large branching fraction, its distinctive experimental 
signature and its relatively simple hadronic dynamics.  However, 
the procedure is general and can be applied to a number of other 
final states.
\prdfig{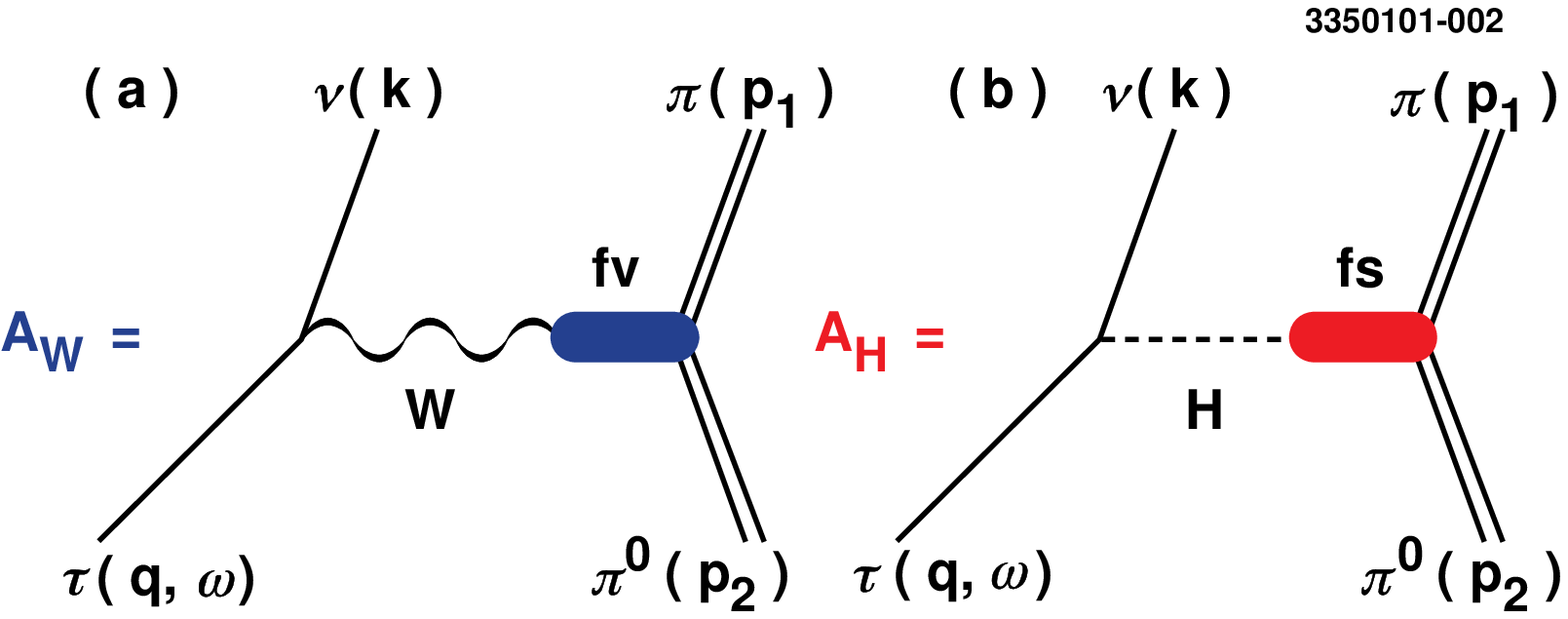}{Aw}{Amplitude for (a) standard W exchange,
(b) scalar exchange.}

The Standard Model amplitude for $\tau$ decay into two pions 
via the W exchange can be written as:
\begin{equation}
\label{eq:Aw}
A_W \sim {\bar{\it{u}}}(\nu)\gamma_{\mu}(1-\gamma_5){\it{u}}(\tau) 
           \underbrace{f_v}_{|f_v|e^{i\delta_v}}Q^{\mu},
\end{equation}
where the hadronic current is parameterized by the relative momentum
between the charged and neutral pions $Q^{\mu} = p_1^{\mu} - p_2^{\mu}$
multiplied by the vector form factor, $f_v$, described by $\rho$ Breit-Wigner
shape:
\begin{equation}
\label{eq:BW}	
 f_v = \frac{-m^2}{s - m^2 + i m \Gamma(s)},
\end{equation}
where $s$ is a squared invariant mass of two pions, $m_{\pi}$ is a pion mass,
and $m$ and $\Gamma(s)$ are the mass and the momentum-dependent width of 
the resonance, respectively. The latter is defined as:
\begin{equation}
\Gamma(s) = \left \{ \begin{array}{ll}
\frac{m}{\sqrt{s}}(\frac{s-4m_{\pi}}{m^2-4m_{\pi}^2})^{3/2}
  & \mbox{if $s>(2m_{\pi})^2$} \\
0 & \mbox{elsewhere.}
\end{array}
\label{eq:BW_width}
\right .
\end{equation}
Here, we neglect the contribution from the $\rho^\prime$ resonance \cite{rhoprime}.

The amplitude for the $\tau$ decay into $\pi^\pm \pi^0$ via a charged Higgs 
boson in the 3HDM model can be written as~\cite{mhdm}:
\begin{equation}
\label{eq:Ah}
A_H \sim {\bar{\it{u}}}(\nu)(1-\gamma_5){\it{u}}(\tau)~
\underbrace{\frac{m_{\tau}}{m_{Higgs}^2}[m_u Z^* X - m_d Z^* Y]}
            _{\Lambda}
           \underbrace{f_s}_{|f_s|e^{i\delta_s}}~ M,
\end{equation}
where $m_{\tau}$, $m_{Higgs}$, $m_u$, and $m_d$ are the masses of the tau 
lepton, charged Higgs, $u$ quark and $d$ quark, respectively.

$X$, $Y$, and $Z$ are the ratios of the complex Higgs couplings to 
$u$, $d$ quarks and leptons relative to the Standard Model weak 
couplings.
The overall Higgs coupling to the $\pi \pi^0$ system is denoted by $\Lambda$:
\begin{equation}
\Lambda = \frac{m_{\tau}}{m_{Higgs}^2}[m_u Z^* X - m_d Z^* Y].
\label{eq:Higgs_lambda}
\end{equation}
Because $\Lambda$ is complex, the CP-odd phase, $\phi$, comes 
from the imaginary part of the coupling constant:
\begin{equation}
 \Lambda = \Re(\Lambda) + i \Im(\Lambda) = |\Lambda|(\cos \phi + i \sin\phi). 
\end{equation}
We parameterize the hadronic current as a product of a 
dimensional quantity, $M=1$ GeV/$c^2$ providing overall normalization, 
and a scalar form factor $f_s \equiv |f_s| e^{i\delta_s}$. Here
$\delta_s$ is a complex strong phase associated with scalar exchange. 

The choice of $f_s$ is not unambiguous. We
study three cases: one with $f_s = 1$, the second with $f_s$
described by the $a_0(980)$ Breit-Wigner shape and the third with
$f_s$ described by $a_0(1450)$ Breit-Wigner shape (see Eq.~\ref{eq:BW})
with a width given by:
\begin{equation}
\Gamma(s) = \left \{ \begin{array}{ll}
\frac{m}{\sqrt{s}}(\frac{s-4m_{\pi}}{m^2-4m_{\pi}^2})^{1/2}
  & \mbox{if $s>(2m_{\pi})^2$} \\
0 & \mbox{elsewhere.}
\end{array}
\right .
\end{equation}

The matrix element for the $\tau\to \pi\pi^0\nu_{\tau}$ decay 
is given by the sum of the $W$ and Higgs exchange processes:
\begin{equation}
 M(\tau^{-}\to \pi^-\pi^0\nu_{\tau}) = A_W + A_H,
\end{equation}
and is given by a sum of terms defined by equations \ref{eq:Aw} and
\ref{eq:Ah}: 
\begin{equation}
  M(\tau^{-}\to \pi^{-}\pi^0\nu_{\tau}) \sim \bar{u}(k)\gamma_{\mu}
                (1-\gamma_5)u(q,s)|f_v|e^{i\delta_v}Q^{\mu} +
                \Lambda ~
              \bar{u}(k)(1+\gamma_5)u(q,s)|f_s|e^{i\delta_s}M, 
\label{eq:feyn}
\end{equation}
where $q$ is the 4-momentum vector of the $\tau$, $s$ is the polarization 
of the $\tau$ and $k$ is the 4-momentum vector of the neutrino.\\
After calculations detailed in Appendix~A the squared matrix element 
takes the form:
\begin{equation}
\label{eq:Mtau_decay}
|{\cal M}|^2_{\tau^\pm \to \pi^\pm\pi^0\nu} \sim G + s^\mu \omega_\mu, 
\end{equation}
where the spin-averaged component of the total width is:
$$G =  2|f_v|^2[2(qQ)(kQ)-(kq)Q^2] + 2|\Lambda|^2|f_s|^2(qk)$$
\begin{equation}
	\label{eq:Mtau_decay2}
     + 4\Re e(\Lambda) |f_v||f_s|\;\cos\delta ~M_{\tau} (Qk) -
    \underline{4\Im (\Lambda^+)|f_v||f_s|\;\sin\delta ~M_{\tau} (Qk)}, 
\end{equation}
and the product of the tau polarization $s^\mu$ and the polarimeter
vector $\omega_{\mu}$ describing the spin-dependent component of the 
decay width is:
$$  \omega_\mu = 
    \{\mp 2 |f_v|^2 M_{\tau}(2Q_{\mu}(kQ)-k_{\mu}Q^2)
    \pm 2|\Lambda|^2 |f_s|^2 M_\tau k_\mu $$
$$  \mp 4\Re e(\Lambda)|f_v||f_s|\;\cos\delta~(Q_\mu(kq)-k_\mu (qQ)) $$ 
$$  \pm \underline{4\Im (\Lambda^+)|f_v||f_s|\;\sin\delta~(Q_\mu(kq)-k_\mu (qQ))} $$   
\begin{equation}
	\label{eq:Mtau_decay3}
  + 4\Re e(\Lambda)|f_v||f_s|\;\sin\delta~
        e_{\mu\alpha\beta\gamma}q^\alpha Q^\beta k^\gamma 
           + \underline{4\Im (\Lambda^+)|f_v||f_s|\;\cos\delta~
        e_{\mu\alpha\beta\gamma}q^\alpha Q^\beta k^\gamma }\}.
\end{equation}
Here the difference between the strong phases for the vector and scalar 
exchanges is denoted as $\delta \equiv \delta_v - \delta_s $.
The parameter $\Lambda^+$ is defined to be equal to $\Lambda$ for 
$\tau^-$ and equal to its complex conjugate, $\Lambda^*$, for $\tau^+$.  
Underlined terms are CP-odd.

\subsection{CP violation in correlated decays of $\tau$ pairs}

For a single $\tau$ decay we do not know the polarization vector and 
consequently can not construct the spin-dependent term. 
However, the situation is different for 
decays of pairs of $\tau$'s produced in $e^+e^-$ annihilations, 
where the parent virtual photon introduces correlations of the $\tau^+$ 
and $\tau^-$ spins. In the following we denote the momenta of the 
particles deriving from $\tau^+$ with an additional bar symbol in 
order to distinguish them from the momenta of the $\tau^{-}$ decay products.
The probability density for the reaction $e^+e^- \to \tau^+\tau^-$ can be 
written as:
\begin{equation}
\label{eq:Ptotal}
	P_{\tau^+\tau^-\to\pi^-\pi^0\nu~\pi^+\pi^0\bar{\nu}} = 
        G\times [(p\bar{q})^2 + (\bar{p}\bar{q})^2 + m_{\tau}(\bar{p}p)]
         \times \bar{G} + \omega_\mu \tilde{C}^{\mu\nu} \bar{\omega_\nu},
\end{equation}
where $p$, $\bar{p}$, $q$, and $\bar{q}$ are momenta of electron, positron, 
$\tau^-$, and $\tau^+$, respectively. 
The polarimeter vectors of the $\tau^-$ and $\tau^+$ are contracted through
the spin-correlation matrix $\tilde{C}^{\alpha\beta}$ (see Appendix~B for
detailed calculations).

The probability density for $\tau$-pair production and the
subsequent decay of each $\tau$ into the $\pi\pi^0$ final state 
consists of all CP-even and odd terms from the 
single $\tau$ decay contracted  
either through the spin-averaged production of the $\tau$ pair 
$(p\bar{q})^2 + (\bar{p}\bar{q})^2 + m_{\tau}(\bar{p}p)$
or through the spin-correlation matrix $\tilde{C}^{\mu\nu}$ with 
CP-even and odd terms of the other $\tau$ decay 
(Eq.~\ref{eq:Mtau_decay}). Due to the spin correlation, these contracted 
terms are no longer proportional to the spins of the individual $\tau$ 
leptons and, therefore, they do not vanish after integration over all 
Lorentz-invariant phase space. 

To design an optimal method for searching for CP violation, we separate 
the CP-even and CP-odd terms in the $\tau$ production and decay 
probability density (Eq.~\ref{eq:Ptotal}). 
We refer to the sum of all CP-even and CP-odd terms as $P_{even}$
and $P_{odd}$, respectively:
\begin{equation}
\label{eq:Ptotal2}
	P = P_{even} + P_{odd}.
\end{equation}
Formulae for $P_{even}$ and $P_{odd}$ are given in Appendix~C.
Both the CP-even and CP-odd parts of the cross section are functions 
of the kinematical quantities that characterize 
the $\tau^+$ and $\tau^-$ decays.

\section{Experimental Method}
\label{ss_method}
\subsection{Optimal variable}
\label{ss_optimal}

A common method used in searches for CP violation is to define 
a CP-odd observable $\xi$, such as, {\sl e.g.}, a triple product of 
independent vectors, and then average its distribution over the data set. 
An average of $\xi$ different from zero indicates the presence of a  
CP-odd component, $P_{odd}(\Lambda)$, of the probability density
which, in general, depends on scalar coupling $\Lambda$. 
The choice of the CP-odd observable is not unique. 
One can always add any CP-odd term to any selected CP-odd observable to 
obtain another CP-odd observable.  Different observables have, in general, 
different sensitivity to CP violation. 
To maximize the sensitivity of our search we construct an optimal variable, 
$\xi$, with the smallest associated statistical error. 
Such a variable was proposed by Atwood and Soni~\cite{optim1} and by 
Gunion and Grzadkowski~\cite{optim2} for other searches. 
The derivation of $\xi$ is described in detail in Appendix~D. The variable
is equal to the ratio of the CP-odd and CP-even parts of the total 
cross section assuming that the absolute value of the coupling 
$\Lambda$ is unity:
\begin{equation}
\label{eq:xi_opt}
	\xi = \frac{P_{odd}(1)}{P_{even}}.
\end{equation}
The average value of $\xi$ is:
\begin{equation}
 <\xi> = \int \frac{P_{odd}(1)}{P_{even}}
               (P_{even} +  P_{odd}(\Lambda)) dLips  = 
          \int \frac{P_{odd}(1) P_{odd}(\Lambda)}{P_{even}} dLips.
\end{equation}
Since $P_{odd}$ is proportional to the imaginary part of $\Lambda$,
we can express it as: 
\begin{equation}
 P_{odd}(\Lambda) = \Im(\Lambda) P_{odd}(1),
\end{equation}
and
\begin{equation}
\label{eq:xi_ave}
 <\xi> = \Im(\Lambda) \int \frac{P_{odd}(1)^2}{P_{even}} dLips. 
\end{equation}
The integral in Eq.~\ref{eq:xi_ave} is always larger than or equal to
zero, and equality occurs only if the odd part of the cross section 
vanishes.  Therefore, we expect that the average value of $\xi$ will be  
proportional to the imaginary part of the Higgs coupling constant and 
will be positive if $\Im(\Lambda) > 0$ and negative if $\Iml < 0$. 
Monte Carlo simulation of the $\xi$ distributions for the three choices 
of scalar form factors and no CP 
violation\footnote{We use a modified version of the TAUOLA $\tau$ decay 
simulation package~\cite{KORALB} to generate Monte Carlo samples 
corresponding to different values of the complex coupling $\Lambda$.} 
are shown in Fig.~\ref{fig:fs}(a).
The same distributions for the CP violating case $\Im(\Lambda)=1$ are
shown in Fig.~\ref{fig:fs}(b).
\prdfig{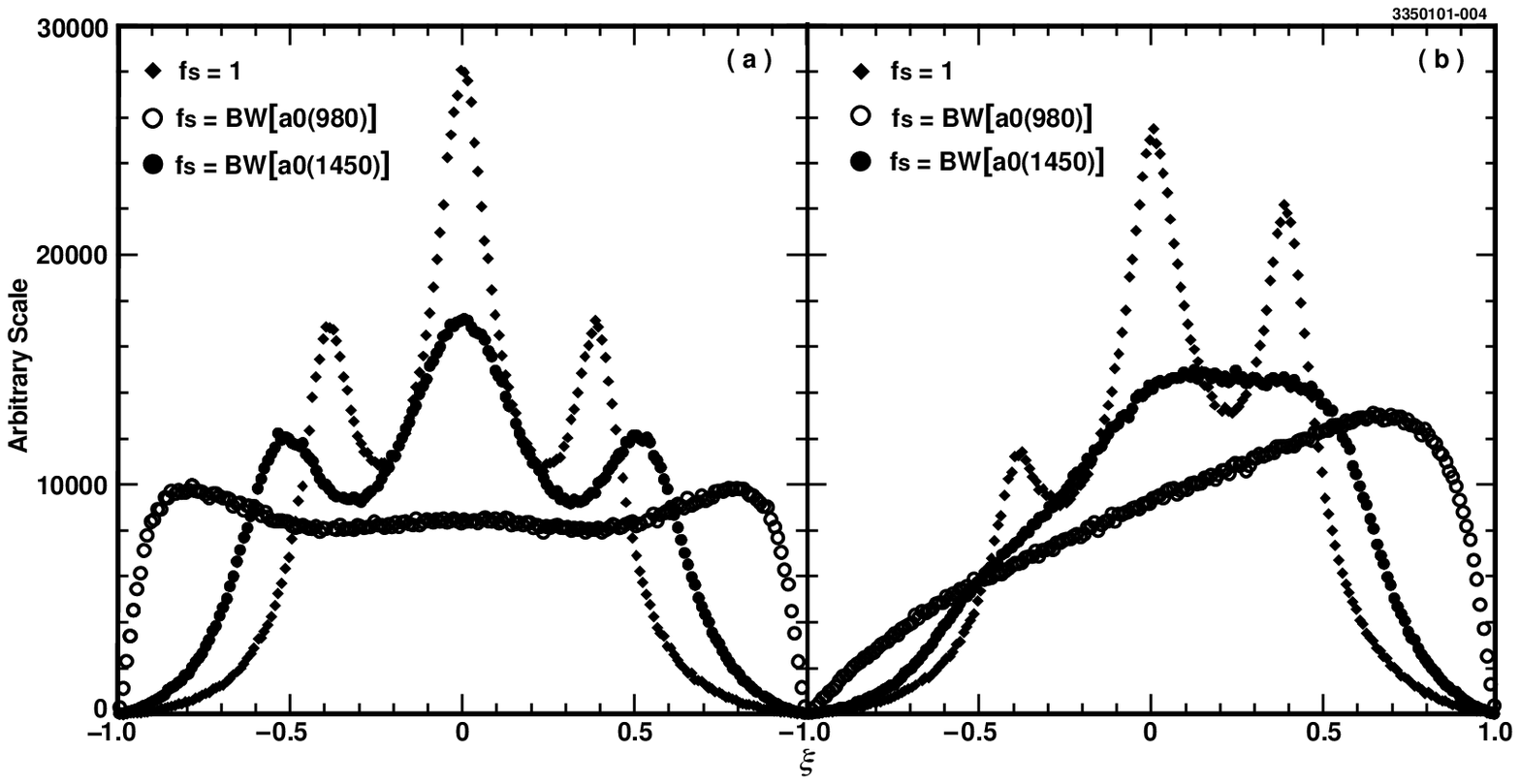}{fs}{Optimal observable, $\xi$,
for (a) Monte Carlo with no CP violation and  (b) Monte Carlo with a maximal CP violation
$\Im(\Lambda)=1$.}
The structure in these distributions is due to the resonant structure in the 
vector and scalar form factors.
%
%
%
\subsection{Experimental issues}
\label{ss_expissues}

To calculate a value of $\xi$ as described in Section~\ref{ss_optimal}, 
we need to know the directions of the outgoing neutrinos, or equivalently 
the directions of the $\tau$'s.  In cases where both $\tau$'s decay to 
semi-hadronic final states with two missing neutrinos we can use energy 
and momentum constraints to determine these directions with a 
two-fold ambiguity~\cite{cone}.
We assume that in the $e^+e^-$ annihilation the two taus are produced 
back to back and we ignore the effects of initial state radiation.
We cannot distinguish the correct solution from 
the incorrect one. We check that the wrong solution
does not introduce any asymmetry (discussed below) and we sum the $\xi$ 
distributions obtained from the two solutions. Finally, we calculate
the mean value of the summed $\xi$ distribution to search for any evidence
of asymmetry.

To assess the possible bias in computed $\xi$ values,  
we use the KORALB/TAUOLA Monte Carlo to calculate the mean value of the 
CP-odd observable for the correct and incorrect solutions separately, 
as well as for the sum of the two.  
Within errors, we observe no artificial asymmetry in the $\xi$ 
distribution for the false solution in the Standard Model Monte Carlo 
sample. 
However, in the sample with CP violation, the asymmetry obtained using 
just the false solution is significantly different from that obtained 
using the correct solution. The Monte Carlo procedure used 
to calibrate this effect is described in Section \ref{results}.
\section{Data Analysis}
\label{analysis}

\subsection{Data and Monte Carlo Samples}
\label{ss_samples}

The data used in this analysis were collected at the Cornell 
Electron Storage Ring (CESR) at or near the energy of the $\Upsilon(4S)$. 
The data correspond to a total integrated luminosity of 13.3 fb$^{-1}$ and 
contain 12.2 million $\tau^+\tau^-$ pairs. Versions of the CLEO detector 
employed here are described in Refs.~\cite{CLEOII} and~\cite{CLEOII5}.
From this data sample, we select events consistent 
with $e^+e^-\to\tau^+\tau^-$ interactions where each 
$\tau$ decays into the $\pi\pi^0\nu_\tau$ final state.  We refer 
to such events as $\rho$-$\rho$ events since this final state   
is dominated by production and decay of the $\rho(770)$ resonance.  
The event selection criteria are summarized in 
Section~\ref{ss_data_selection}.  

To estimate backgrounds we analyze large samples of Monte Carlo events 
following the same procedures that are applied to the actual CLEO data.  
The physics of $\tau$-pair production and decay is modeled by the 
{\rm KORALB/TAUOLA} event generator~\cite{KORALB}, while the detector 
response is handled with a GEANT-based~\cite{GEANT} simulation of the 
CLEO detectors.
For backgrounds coming from $\tau$ decays other than 
$\tau\to\pi\pi^0\nu_{\tau}$, we analyze a Monte Carlo sample containing 
37.2 million $\tau^+\tau^-$ events in which all combinations of 
$\tau^+$ and $\tau^-$ decay modes are present, except for our 
signal $\rho$-$\rho$ process. 

Non-$\tau$ background processes include annihilation into 
multihadronic final states, namely $e^+e^-\to q\overline{q}$ 
($q = u,\,d,\,s,\,c$ quarks) and $e^+e^-\to\Upsilon(4S)\to B\overline{B}$, 
as well as production of hadronic final states due to two-photon 
interactions.  
Backgrounds from multihadronic physics are estimated using 
Monte Carlo samples which are slightly larger than the CLEO data 
and contain 42.6 million $q\overline{q}$ 
and 17.3 million $B\overline{B}$ events, respectively. 
The background due to two-photon processes is estimated 
from Monte Carlo simulation of 37,556 $2\gamma\to\tau^+\tau^-$ events, 
using the formalism of Budnev $et~al.$\cite{gamgam}. 
All non-tau backgrounds are found to be negligible (see Section~IV.C).
\par
To study the potential bias of the mean value
of the observable $\xi$ (Eq.~\ref{eq:xi_opt}) that can be introduced by
data selection, we use Monte Carlo samples generated with and without 
CP violation.  
The ``Standard Model'' sample (no CP violation) consists of  
2.4~million $\rho$-$\rho$ events.  
The equivalent integrated luminosity of this Monte Carlo samples  
corresponds to four times that of the CLEO data set.
Each sample of the CP-violating Monte Carlo contains 1.5 million events. 

\subsection{Event Selection.}
\label{ss_data_selection}

Tau leptons are produced in pairs in $e^+e^-$ collisions. 
The event selection follows mostly the procedure developed originally 
for the study of the $\tau\to\rho\nu_{\tau}$ decays \cite{rho-rho}.
At CESR beam energies, the decay products of the $\tau^+$ and 
$\tau^-$ are well separated in the CLEO detector.  We require an event 
to contain exactly two reconstructed charged tracks, 
separated in angle by at least $90^0$. 
The net charge of the event must be equal to zero. Both
tracks must be consistent with originating from $e^+e^-$ interaction
region to reject events arising from beam-gas interactions 
or from cosmic muons passing through the detector.   
Each track must have a momentum smaller than 
0.85 $E_{beam}$ to minimize background from Bhabha scattering and muon 
pair production. The momenta of all charged tracks are corrected 
for dE/dx energy loss in the beam pipe and tracking system. 
No attempt is made to distinguish $\pi^\pm$ from charged leptons or  
$K^\pm$ mesons: all charged particles are assumed to be pions. 

Clusters of energy deposition in the calorimeter are considered as
candidates for photons from $\pi^0$ decay if they are observed in the 
central part of the detector ($|\cos\theta|<0.707$, where $\theta$ is 
the angle between the positron beam axis and the photon direction), 
are not matched to a charged track, and have energy greater than 30 MeV. 
A cluster must have
a transverse energy profile consistent with expectations for a photon,
and it must be at least 30 cm away from the nearest track projection.
We select events with only four photon candidates. 
Pairs of photons with a reconstructed invariant mass $M_{\gamma\gamma}$ 
between $-4$ and $+3$ standard deviations ($\sigma_{\gamma\gamma}$) of the 
$\pi^0$ mass are considered to be $\pi^0$ candidates. 
The $\pi^0$ invariant mass resolution, $\sigma_{\gamma\gamma}$, 
varies from 4 to 7 MeV/$c^2$, depending on the $\pi^0$ energy and decay angle. 
To reject events with spurious $\pi^0$ candidates resulting from 
random combinations of low energy clusters, the $\pi^0$ energy is 
required to be greater than 250 MeV (see Fig.~\ref{fig:fakepi0}). 
\prdfig{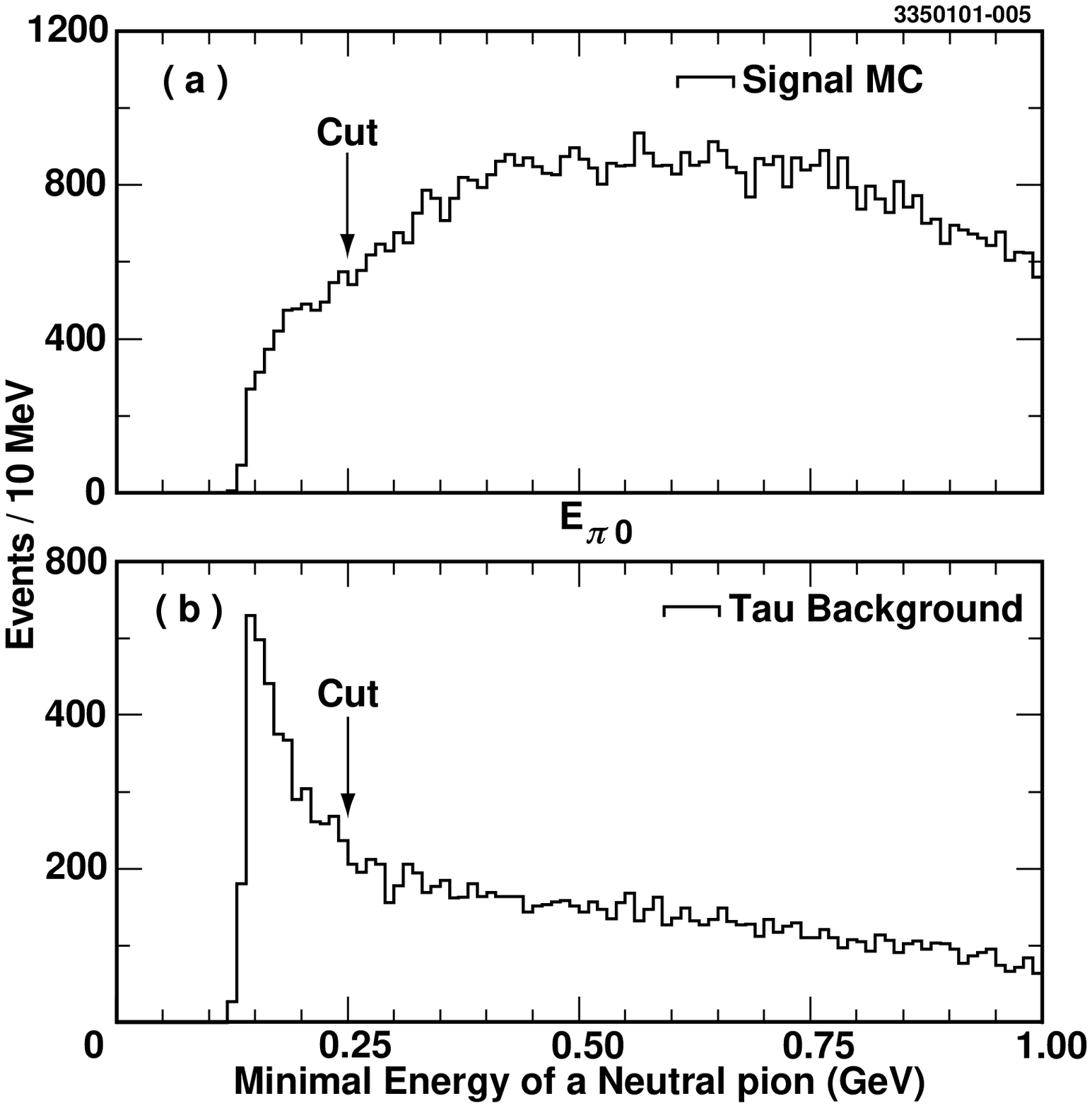}{fakepi0}{Energy of the neutral
pions: (a) signal Monte Carlo, (b) tau background.}
Each $\pi^0$ candidate is associated with the nearest 
in angle charged track to form a $\pi\pi^0$ candidate.  
Monte Carlo studies show that this method of assignment is 
correct 99\% of the time. 

The event selection criteria successfully rejects background from
high multiplicity multihadronic events ($e^+e^- \to q\bar{q}$).
Background from low-multiplicity two-photon ($e^+e^-\to e^+e^-\gamma\gamma$) 
events are rejected by requiring that the missing transverse momentum 
be larger than 200 MeV/c and that the missing mass of the
event to be smaller than 7.5 GeV/$c^2$ (see Fig.~\ref{fig:2gamma}).
Overall, data selection efficiency is estimated to be 7.4\%.
Extensive systematic studies of the background dependence on selection
criteria cuts have been done in Ref.~\cite{rho-rho} and confirmed 
by the present work.
\prdfig{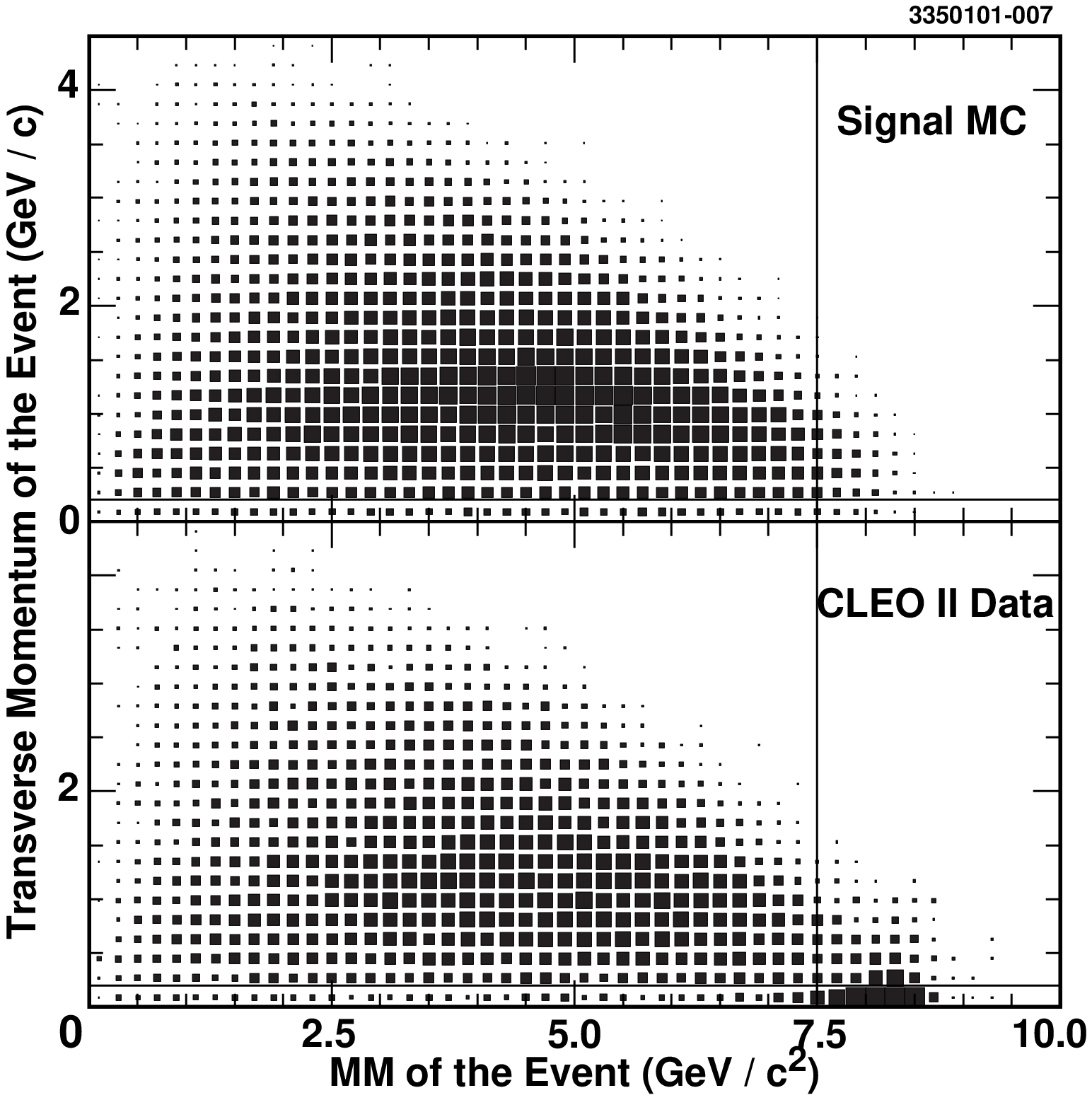}{2gamma}{Transverse momentum and
missing mass of the event for the data and signal Monte Carlo.
The vertical line indicates the position of the cut.}
%
%
%
\subsection{Estimate of the remaining background}
\label{ss_background}

After applying the same selection criteria to the Monte Carlo simulation of 
$e^+e^-\to q\bar{q}$ and $e^+e^- \to e^+e^-\gamma\gamma$ processes, 
we estimated that the remaining background from those sources represents 
less than 0.1\% of the selected data sample.
Consequently we ignore this background in our  
extraction of $<\xi>$.   We also find no evidence for contribution of cosmic 
ray or beam-gas interactions in the selected data.
The main remaining background is due to the $\tau$-pair events in which one of the 
$\tau$'s decays into $\rho\nu_{\tau}$ while the other decays into 
$\pi 2\pi^0\nu_{\tau}$ and the photons from one of the $\pi^0$'s are not detected. 
The contamination from this background source is estimated to be 5.2\%. 
The second largest background contribution of 2.1\% is due to one of 
the $\tau$'s decaying into the $K^*\nu_\tau$ final state producing 
a neutral pion plus a charged kaon which is mistaken 
for a pion. All other tau decays provide much
smaller contributions with the largest being less than 0.7\%. The total 
background contamination from tau decays is estimated to be 9.9\%. 
%
%
\section{Results}
\label{results}
\subsection{Calibration of the parameterization after applying selection criteria}
\label{ss_calibration}
To relate the observed mean value of the optimal observable, $<\xi>$, to
the imaginary part of the coupling constant $\Lambda$, the $\Im(\Lambda)$
dependence of $<\xi>$ must be known. To simplify the notation, 
in the following we define $\lambda$ to be the imaginary part 
of the scalar coupling constant $\Lambda$. 
\par
The CP-even terms for each single tau decay are either independent 
of $\lambda$ or proportional to $\lambda^2$, while CP-odd terms are 
either linear functions of $\lambda$ or proportional to $\lambda^3$.  
Therefore, $\xi$, which is a product of CP-even and CP-odd terms, must
have only odd power terms in $\lambda$ expansion and to the first 
order the mean value is proportional to $\lambda$ with a proportionality 
coefficient $c$:
\begin{equation}
 <\xi> = c \times \lambda.
\label{eq:xi_powers}
\end{equation}
To check this assertion and to calculate the regions of $\lambda$
where the linear dependence holds we generated 21 
event-generator-level Monte Carlo samples (no detector simulation) 
with different values of $\lambda$ varying between -1 and 1, 
and calculated the mean value  $<\xi>$ for each one.  
Each sample comprised $10^6$ events. 
For each form of the scalar component, the calculated asymmetry distribution 
was then fit to a fifth order polynomial (see Fig.~\ref{fig:fit_lin}).
Using these distribution we find the ranges of $\lambda$ for which
$<\xi>$ is linear in $\lambda$.
As an additional cross-check of our method, we search for (and 
find no evidence of) terms proportional to even powers of $\lambda$. 
\prdfig{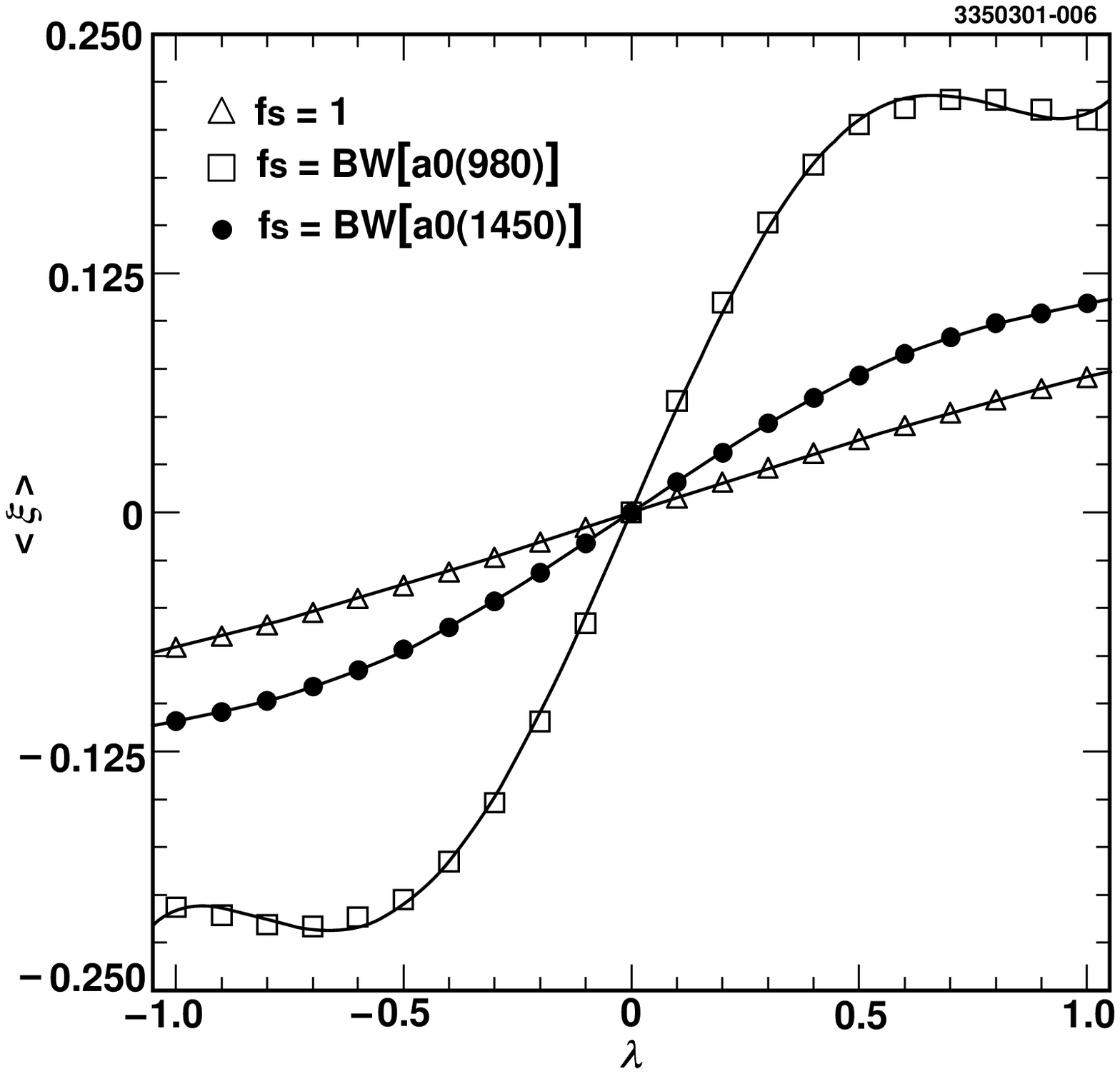}{fit_lin}{$\lambda$ dependence on
average value of optimal observable $<\xi>$, for different assumptions
on scalar form factor $f_s$.}
\par
Measurement errors can introduce a bias in the $\xi$ 
distribution (see Eq.~\ref{eq:xi_opt}) as well as change
the linearity coefficient $c$ (Eq.~\ref{eq:xi_powers}) due to event
selection, reconstruction and resolution. Therefore, this coefficient 
is calculated after applying selection criteria to events generated with 
full GEANT-based detector simulation and pattern recognition software. 
For each choice of the scalar form factor $f_s$
we use five samples of 200,000 signal Monte Carlo events generated with 
different values of $\lambda$ in the region where $<\xi>$ is
linear in $\lambda$. For each sample, we calculate the average value 
of the optimal observable and plot it as a function of $\lambda$.
For each form of the scalar component, the calculated asymmetry distribution 
is fit to a straight line to obtain the calibration coefficients.
To check that the selection criteria do not create an artificial 
asymmetry we calculate the
mean value of the optimal observable for Standard Model Monte Carlo
samples for each choice of the scalar form factor. We list these values along
with calculated coefficients $c$ in Table~\ref{tab:coeff}.
\begin{table}[htb]
\begin{center}
\caption{\label{tab:coeff} 
Average values of the optimal observable $<\xi>$ for the Standard Model 
Monte Carlo and the proportionality coefficient $c$ for the CP asymmetry 
fits for different scalar form factors.}
\begin{tabular}{|c|c|c|}
Form factor, $f_s$       & $<\xi>$, $10^{-3}$     & $c$, $10^{-3}$ \\ \hline
1                       & $ 0.7\pm0.6$           & $66.8\pm4.3$   \\ 
$BW(a_0(980))$          & $ 1.0\pm1.1$           & $586.4\pm19.4$ \\ 
$BW(a_0(1450))$         & $ 0.5\pm0.8$           & $145.8\pm7.3$  \\
\end{tabular}
\end{center}
\end{table}
\par
We observe that the event selection criteria do not introduce 
an artificial asymmetry in the $\xi$ distribution. For all three
form factors, the mean value of $\xi$ for the Standard Model Monte 
Carlo sample is consistent with zero within its statistical error. 
This fits show no indication that $<\xi>$ deviates from linearity
within the range determined by event-generator-level Monte Carlo study.
We then use the coefficients from Table~\ref{tab:coeff} to 
calculate $\lambda$. 
%
\subsection{Observed mean values}
\label{ss_results}
For each choice of the scalar form factor, we obtain a distribution
in $\xi$, with two entries per event for the two solutions for the 
tau lepton direction as described in Section III. These distributions
are shown in Fig.~7, with those from the Standard Model Monte Carlo
simulations overlaid. From these distributions we compute the mean values
$<\xi>$ after subtracting the average value for Standard Model Monte Carlo,
which are reported in the first column of Table II. In each case,
we use the appropriate empirically-determined coefficient given in
Table I to derive a value for the imaginary part of the Higgs coupling 
$\Lambda$, as described in the preceding section. These values along with
the $90\%$ confidence limits on $\Im(\Lambda)$ are reported in the second
and third columns of Table II.

\prdfig{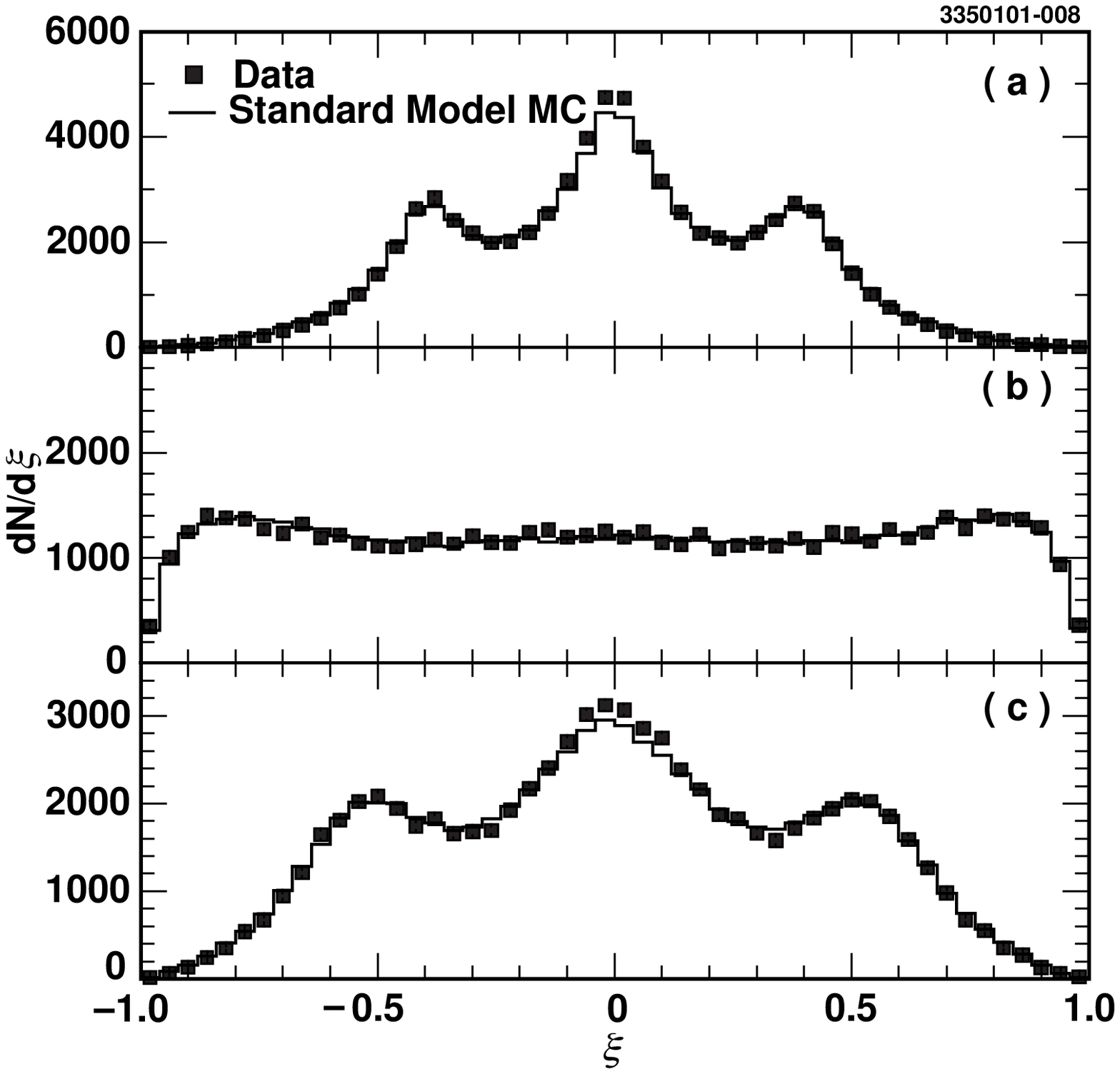}{paper_asym}{The distribution of 
CP violation sensitive variable $\xi$ for the data (dots) compared to 
the Standard Model Monte Carlo prediction (solid line) for (a) $f_s$=1, 
(b) $f_s=BW(a_0(980))$ and (c) $f_s=BW(a_0(1450))$.}
\begin{table}[htb]
\caption{\label{tab:results} 
Average value of the optimal observable $\xi$ after subtracting the 
average value for Standard Model Monte Carlo, calculated value of 
$\Im(\Lambda)$ and 90\% C.L. limits on $\Im(\lambda)$.}
\begin{center}
\begin{tabular}{|c|c|c|c|}
Form factor                & $<\xi>$, $10^{-3}$ & $\Im(\Lambda)$, $10^{-2}$ 
                                                       & $\Im(\Lambda)$, 90\% confidence limits \\ \hline
$f_s = 1$                 & $-0.8\pm1.4$       & $-1.2\pm2.1$         & $-0.046 < \lambda < 0.022$ \\
$f_s = BW(a_0(980))$      & $-0.6\pm2.4$       & $-0.1\pm0.4$         & $-0.008 < \lambda < 0.006$ \\
$f_s = BW(a_0(1450))$     & $ 0.2\pm1.7$       & $ 0.1\pm1.2$         & $-0.019 < \lambda < 0.021$ \\
\end{tabular}
\end{center}
\end{table}
%
%
\subsection{Systematic errors}
\label{ss_syserr}

In principle, several possible sources of systematic error 
can contribute to this analysis.  We found no sizable effect 
to alter the limits shown in Table \ref{tab:results}.
We describe most significant effects below.

\subsubsection{Detector asymmetry}
The detector can create an artificial CP asymmetry due to the 
imperfect tracking, detection efficiency, and resolution.
To check the symmetry of the CLEO detector tracking,
an independent analysis was done using leptonic tau decays. 
Such decays have been studied with $\sim0.1\%$ precision \cite{PDG_leptonic}
and at that level show no deviation from the Standard Model.
Therefore, any observed asymmetry would indicate detector
effects. We find the detector response symmetric with respect
to the charge within a precision of $0.2\%$ and 
the detector asymmetry, $A$, in momentum distribution is:
$$ -0.4\% < A < 0.2\% \mbox{ at 90\% C.L.}$$

\subsubsection{Track reconstruction efficiency}
We study the systematic effects due to a possible difference 
of the track reconstruction efficiency for $\pi^-$ and $\pi^+$ 
as a function of the pion momentum. To estimate the size of this 
effect we plot in  Fig.~\ref{fig:pipi0_syst}(a) the momentum
distribution for charged pions in the reaction 
$\tau^\pm \to \pi^\pm \pi^0 \nu_{\tau}$. The ratio of these distributions
shown in Fig.~\ref{fig:pipi0_syst}(b) is consistent with 1.
The introduction of a $\pm 1\sigma$ slope to the ratio plotted
in Fig.~\ref{fig:pipi0_syst}(b) does not change the value of the
coefficient $c$ but changes the value of $\Im(\Lambda)$ by $\pm0.003$.
We take this as a measure of a systematic error.
\prdfig{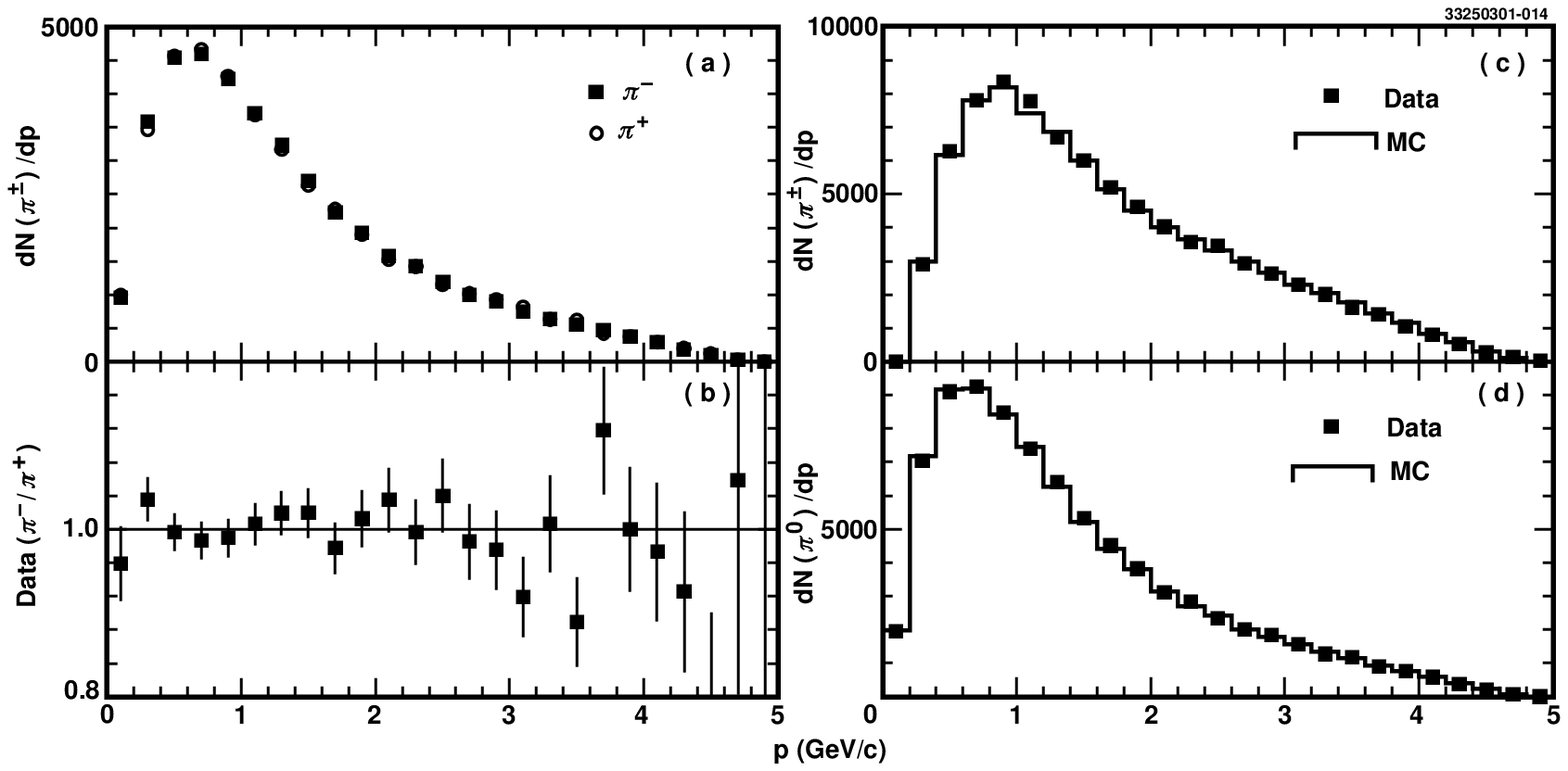}{pipi0_syst}{ Momentum distributions for (a) $\pi^-$ and $\pi^+$ in data, 
(b) $\pi^-$/$\pi^+$ ratio in data, (c) charged pions, (d) neutral pions.}

\subsubsection{Momentum reconstruction}
Another possible source of systematic error is due to an imperfect Monte
Carlo simulation of the momentum distributions for the tau decay products.
Monte Carlo describes the data very well [see Fig.~\ref{fig:pipi0_syst}(c) and
Fig.~\ref{fig:pipi0_syst}(d)]. We estimate the size of the systematic error
by introducing artificial $\pm 1\sigma$ shifts in the slopes of the
charged and neutral pions momentum distributions. These shifts do not bias
the $<\xi>$ measurement, but change the calibration coefficient $c$. 
The systematic error on $c$ is estimated to be $2.2\times10^{-3}$
and has a negligible effect ($\sim 10^{-5}$) on the value of $\Im(\Lambda)$.

\subsubsection{Background}
Another source of possible distortion of the result is an asymmetry
of the $\xi$ distribution induced by the remaining background. 
In order to estimate this effect we denote by $\xi_S$ and $\xi_B$
the values of $\xi$ from the signal and background events, respectively.
If we denote the number of signal and background events by
$S$ and $B$, then, the observed average value of $\xi$ is:
\begin{equation}
	<\xi> = \frac{\sum \xi_S + \sum \xi_B}{S + B} =
                \frac{\sum \xi_S}{S}\times\frac{S}{S+B} + 
                \frac{\sum \xi_B}{B}\times\frac{B}{S+B}.
\end{equation}
Since the number of background events in this analysis is much smaller
than the number of signal events, then this equation can be simplified:
\begin{equation}
	<\xi> \sim \frac{\sum \xi_S}{S+B} + \frac{\sum \xi_B}{B}
                                       \times\frac{B}{S} 
               = <\xi_S> + <\xi_B>\times \frac{B}{S},
\end{equation}
where $<\xi_S>$ is a mean value due to signal and $<\xi_B>$ is a 
mean value due to background. We estimate $<\xi_B>B/S$ using Monte Carlo:
$$<\xi_B>\times\frac{B}{S} =(0.04\pm0.04)\times10^{-3}.$$
Thus, the systematic uncertainty arising from this source is negligible on
the scale of the sensitivity of our measurement.

\subsection{Summary}
\label{ss_summary}
Within our experimental precision we observe no significant 
asymmetry of the optimal variable and, therefore, no CP violation in 
$\tau$ decays. Due to the uncertainty in the choice of the scalar form factor
we select the most conservative 90\% confidence limits corresponding to 
$f_s=1$:
$$ -0.046 <\lambda < 0.022, \mbox{ at 90\% C.L.}. $$
These limits include the effects of possible systematic errors.
%
%
\section{\boldmath Search for scalar-mediated $\tau$ decays}
\label{helicity}

The helicity angle, $\theta_{\pi\pi}$, is defined as  
the angle between the direction of the charged pion in the $\pi\pi^0$ 
rest frame and the direction of the $\pi\pi^0$ system in the
$\tau$ rest frame. In Standard Model, the helicity angle is expected 
to have a distribution corresponding to a vector exchange:
\begin{equation}
 \frac{dN}{d~cos\theta_{\pi\pi}} \sim a + b\cos^2\theta_{\pi\pi}.
 \label{eq:helDIS}
\end{equation}
For scalar-mediated decays, there is an additional term proportional 
to $\cos\theta_{\pi\pi}$ that corresponds to the S-P wave interference 
and linearly proportional to the scalar coupling constant $\Lambda$.
In general, $\Lambda$ is complex and the term linear in $\cos\theta_{\pi\pi}$
is proportional to the real and imaginary parts of the scalar coupling 
with coefficients $c_1$ and $c_2$, respectively:
\begin{equation}
 \frac{dN}{d~cos\theta_{\pi\pi}} \sim a + 
                     c_1 \Re(\Lambda) \cos\theta_{\pi\pi}
                   + c_2 \Im(\Lambda) \cos\theta_{\pi\pi}
                                     + b\cos^2\theta_{\pi\pi}.
 \label{eq:helDIS2}
\end{equation}
The observation of the terms proportional to cosine of the helicity
angle would indicate the scalar exchange in the tau decays. In the following,
we discuss a model independent method used to extract confidence limits on 
the imaginary part of the scalar coupling $\Lambda$.
\par
In order to calculate the helicity angle we must know the tau rest frame.
Due to the unobserved neutrino, the tau rest frame can only be reconstructed
with a two-fold ambiguity. We can avoid such ambiguity by using the 
pseudo-helicity angle, $\theta^*$. This pseudo-helicity angle is obtained
by replacing the tau rest frame with the laboratory rest frame where it 
is defined as an angle between the direction of $\pi^\pm$ in the $\pi\pi^0$ 
rest frame and the direction of the $\pi\pi^0$ system in the lab frame.
The difference between the pseudo-helicity distributions for the
$\tau^+$ and $\tau^-$ decays is expected to have the same form as 
given by Eq.~\ref{eq:helDIS2} but with a different numerical coefficients.
\par
The term including $\Im(\Lambda)$ changes sign for
tau leptons of opposite charges. Therefore, the difference of the 
pseudo-helicity distributions for positive and negative tau leptons 
has the term linear in $\cos\theta^*$ proportional to the imaginary part 
of the scalar coupling $\Lambda$ only:
\begin{equation}
 \frac{dN(\tau^-)}{d~cos\theta^*} -  \frac{dN(\tau^+)}{d~cos\theta^*}
 \sim 2 c_2 \Im(\Lambda) \cos\theta^*.
 \label{eq:hel_observable2}
\end{equation}
The presence of this term indicates CP violation.
\par
In this study, we use the same data sample as for the previous analysis
with the same selection criteria except for the requirement of the 
successful cone reconstruction. The pseudo-helicity distribution
for $\tau^-$ and $\tau^+$ is given in Fig.~\ref{fig:hel_data}(a).
\par
The structure in Fig.~\ref{fig:hel_data}(a) is due to the variation in the 
efficiency as a function of charged pion momentum and $\pi^0$ energy.
To obtain the product of the imaginary part of the scalar 
coupling $\Lambda$ and a linearity coefficient $c_2$ 
(see Eq.~\ref{eq:hel_observable2}), we fit the difference 
of the two pseudo-helicity distributions 
for negative and positive tau leptons to a first order polynomial. 
To minimize systematic effects due to soft
pion reconstruction we perform the fit in the region of 
$-0.7<\cos\theta^*<0.8$, which
corresponds to pions with momentum higher than 0.3 GeV/$c$.
The obtained value of the slope for the data distribution is:
$$ c_2 \Im(\Lambda) = (4.2 \pm 3.6) \times 10^{-4}$$
and illustrated in Fig.~\ref{fig:hel_data}(b).
It is consistent with zero within statistical error.
\prdfig{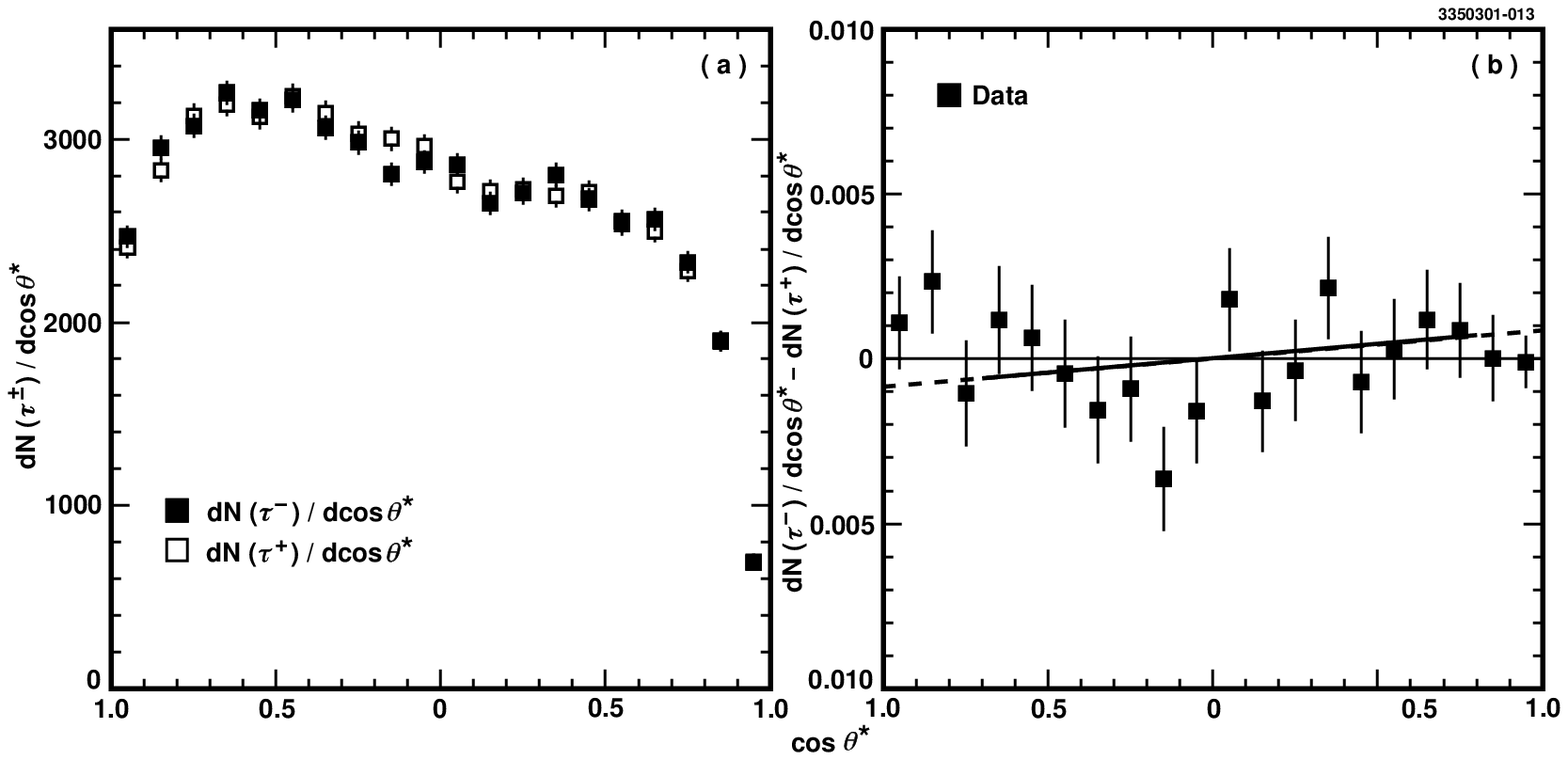}{hel_data}{(a) Pseudo-helicity
distribution for $\tau^-$ and $\tau^+$ in data, (b) difference
between pseudo-helicity distributions for $\tau^-$ and $\tau^+$ in data.
Solid line is a linear fit and dashed lines show extrapolation to the
region excluded from the fit.}
\par
To determine the proportionality coefficient $c_2$ we follow
the procedure described in Section V. We calculate the slope
of the difference of pseudo-helicity distributions
after applying selection criteria  using five samples
of 200,000 signal Monte Carlo events generated with different values
of $\lambda$. Then, we fit the calculated slope dependence
as a function of $\lambda$ using straight line to obtain $c_2$:
$$ c_2 = (107.5 \pm 12.6)\times 10^{-4}.$$
We use this coefficient to obtain the value of the imaginary part of the
scalar coupling $\Im(\Lambda)$:
\begin{equation}
\Im(\Lambda) = 0.028 \pm 0.037, \mbox{ or } -0.033 < \Im(\Lambda) < 0.089
   \mbox{ at 90\% C.L.}
\label{eq:hel_result_im}
\end{equation}
As expected, the limit on the $\Im(\Lambda)$ is less strict than 
the one obtained using the optimal observable.
%
%
%
%
\section{Conclusions}
\label{conclusion}
We have discussed a method of searching for the CP violation
in the correlated decays of two tau leptons each decaying 
into the $\pi^\pm\pi^0\nu_{\tau}$ final state. The limit on the imaginary 
part of Higgs coupling constant is model dependent and depends on 
the choice of scalar form factor. The most conservative choice of 
scalar form factor $f_s=1$ leads to a conservative limit on a scalar coupling:
$$ -0.046 < \Im(\Lambda) < 0.022 \mbox{ at 90\% C.L.}$$
Using pseudo-helicity method we obtain a limit on a
imaginary part of the scalar component in the tau decays:
$$ -0.033 < \Im(\Lambda) < 0.089 \mbox{ at 90\% C.L.}$$
Both limits agree with each other and restrict the size of the 
contribution of multi-Higgs doublet model diagrams to the $\tau$ lepton decay.
\par
We can relate the limit on $\Im(\Lambda)$ to the limit on the product 
of multi-Higgs model coupling constants (see Eq.~\ref{eq:Higgs_lambda}):
$$  \mbox{-0.026 (GeV/}c^2~)^{-1} < \frac{1}{m_{Higgs}^2}(m_u \Im (Z^*Y) - m_d \Im (Z^*X)) <
    \mbox{ 0.012 (GeV/}c^2~)^{-1}. $$
\section{Acknowledgments}
\label{acknowledgments}
We gratefully acknowledge the effort of the CESR staff in providing us with
excellent luminosity and running conditions.
M. Selen thanks the PFF program of the NSF and the Research Corporation, 
and A.H. Mahmood thanks the Texas Advanced Research Program.
This work was supported by the National Science Foundation, the
U.S. Department of Energy, and the Natural Sciences and Engineering Research 
Council of Canada.
\appendix
\section{Calculation of the squared matrix element 
         for the decay $\tau^-\to \pi^-\pi^0\nu_\tau$}

The matrix elements and their conjugates have the following forms:
\begin{equation}
 M(\tau^{-}\to \pi^{-}\pi^0\nu_{\tau}) \sim 
    f_{V}Q^{\mu}\bar{u}(k)\gamma_{\mu}(1-\gamma_5)u(q,s) +
   \Lambda f_{S}\bar{u}(k)(1+\gamma_5)u(q,s), \label{eq:me1}
\end{equation}
\vspace*{-0.7cm}
\begin{equation}
 M^{*}(\tau^{-}\to \pi^{-}\pi^0\nu_{\tau}) \sim 
    f_{V}^{*}Q^{\mu}\bar{u}(q,s)\gamma_{\mu}(1-\gamma_5)u(k) +
   \Lambda^{*}f_{S}^{*}\bar{u}(q,s)(1-\gamma_5)u(k), \label{eq:me2}
\end{equation}
\vspace*{-0.5cm}
\begin{equation}
 \bar{M}(\tau^{+}\to \pi^{+}\pi^0\nu_{\tau}) \sim 
   -f_{V}Q^{\mu}\bar{v}(q,s)\gamma_{\mu}(1-\gamma_5)v(k) +
   \Lambda^{*}f_{S}\bar{v}(q,s)(1-\gamma_5)v(k), \label{eq:me3}
\end{equation}
\vspace*{-0.5cm}
\begin{equation}
 \bar{M}^{*}(\tau^{+}\to \pi^{+}\pi^0\nu_{\tau}) \sim 
   -f_{V}^{*}Q^{\mu}\bar{v}(k)\gamma_{\mu}(1-\gamma_5)v(q,s) +
   \Lambda f_{S}^{*}\bar{v}(k)(1+\gamma_5)v(q,s). \label{eq:me4}
\end{equation}
The momentum vectors $q$, $K$, $Q$, and form factors $f_v$ and $f_s$ are 
defined in section~\ref{theory}.
We used the following rules to form complex conjugates: we 
conjugate all complex numbers, $u \leftrightarrow \bar{u}$, 
$v \leftrightarrow \bar{v}$, and they exchange places; $1 \leftrightarrow 1$,
$\gamma_{5} \leftrightarrow -\gamma_{5}$, $\gamma_{\mu} \leftrightarrow 
\gamma_{\mu}$, 
$\gamma_{\mu}\gamma_{5} \leftrightarrow \gamma_{\mu}\gamma_{5}$. 
The minus sign at the vector hadronic current of the equations 
(\ref{eq:me3}) and (\ref{eq:me4}) is due to the anti-symmetric behavior 
of the $\rho$ under the isospin rotation.
To form the squared matrix element $|M|^2 = M^{*}M$, we use: 
\begin{equation}
u(\nu)\bar{u}(\nu) = k^{\rho}\gamma_{\rho} = v(\nu)\bar{v}(\nu),
\label{eq:spinors1}
\end{equation}
\begin{equation}
u(\tau)\bar{u}(\tau) = (q^{\alpha}\gamma_{\alpha} + m)(1 + s^{\beta}
\gamma_{\beta}\gamma_{5}),
\label{eq:spinors2}
\end{equation}
\begin{equation}
v(\tau)\bar{v}(\tau) = (q^{\alpha}\gamma_{\alpha} - m)(1 + s^{\beta}
\gamma_{\beta}\gamma_{5}).
\label{eq:spinors3}
\end{equation}
where $s^{\beta}$ is the spin vector(axial) of the tau.
Multiplying matrix elements by their complex conjugates gives us the
absolute value of the squared matrix element:
\begin{eqnarray*}
|M|^2 & = M^{*}M \sim & |f_{V}|^2Q^{\mu}Q^{\nu}(q^{\alpha}\gamma_{\alpha} + m)
                  (1 + s^{\beta}\gamma_{\beta}\gamma_5)\gamma_{\mu}
                  (1 - \gamma_{5})k^{\rho}\gamma_{\rho}\gamma_{\nu}
		  (1 - \gamma_{5})
                     \\
                    && +  |\Lambda|^2|f_{S}|^2(q^{\alpha}\gamma_{\alpha} + m)
                  (1 + s^{\beta}\gamma_{\beta}\gamma_5)
                  (1-\gamma_5)k^{\rho}\gamma_{\rho}
                  (1+\gamma_5) \\
                    && +  \Lambda f_{S}f_{V}^{*}Q^{\mu}
                    (q^{\alpha}\gamma_{\alpha} + m)
                    (1 + s^{\beta}\gamma_{\beta}\gamma_5)\gamma_{\mu}
                    (1 - \gamma_{5})k^{\rho}\gamma_{\rho}
                    (1+\gamma_5)
\end{eqnarray*}
\vspace*{-0.5in}
\begin{equation}
~~~~~~~~~~~~~~~~~~~~~~~~~~~~
                     +  \Lambda^{*}f_{S}^{*}f_{V}Q^{\mu}
                    (q^{\alpha}\gamma_{\alpha} + m)
                    (1 + s^{\beta}\gamma_{\beta}\gamma_5)
                    (1-\gamma_5)k^{\rho}\gamma_{\rho}\gamma_{\mu}
                    (1-\gamma_{5}), \label{eq:A8}
\end{equation}
\begin{eqnarray*}
|\bar{M}|^2 & = \bar{M}^{*}\bar{M}
                \sim & |f_{V}|^2Q^{\mu}Q^{\nu}(q^{\alpha}\gamma_{\alpha} - m)
                  (1 + s^{\beta}\gamma_{\beta}\gamma_5)\gamma_{\mu}
                  (1 - \gamma_{5})k^{\rho}\gamma_{\rho}\gamma_{\nu}
                  (1 - \gamma_5)~~~~~~~~\\
            && +  |\Lambda|^2|f_{S}|^2(q^{\alpha}\gamma_{\alpha} - m)
                  (1 + s^{\beta}\gamma_{\beta}\gamma_5)
                  (1-\gamma_5)k^{\rho}\gamma_{\rho}
                  (1+\gamma_5)\\
            && -  \Lambda^* f_{S}^{*}f_{V}Q^{\mu}
                    (q^{\alpha}\gamma_{\alpha} - m)
                    (1 + s^{\beta}\gamma_{\beta}\gamma_5)
                    (1 - \gamma_{5})k^{\rho}\gamma_{\rho}\gamma_{\mu}
                    (1 - \gamma_5)
\end{eqnarray*}
\vspace*{-0.4in}
\begin{equation}
~~~~~~~~~~~~~~~~~~~
             -  \Lambda f_{S}f_{V}^{*}Q^{\mu}
                    (q^{\alpha}\gamma_{\alpha} - m)
                    (1 + s^{\beta}\gamma_{\beta}\gamma_5)
                    \gamma_{\mu}(1 -\gamma_5)k^{\rho}\gamma_{\rho}
                    (1 +\gamma_{5}). \label{eq:A9}
\end{equation}
Reducing to the traces, we can write the matrix element squared as follows:
\begin{eqnarray*}
|M^{2}| & \sim & 2|f_{V}|^2(2(q \cdot Q)(k \cdot Q)-(q \cdot k)Q^2)
           + 2|\Lambda|^2|f_{S}|^2(q \cdot k)
           + 4\Re(\Lambda f_{S}f_{V}^{*})m(Q \cdot k)\\ 
  && +2|f_{V}|^2m(2(s \cdot Q)(k \cdot Q) - (s\cdot k)Q^2)
  -2|\Lambda|^2|f_{S}|^2m(s \cdot k) \\
  &&  +4\Re(\Lambda f_{S}f_{V}^{*})((s \cdot Q)(q \cdot k) 
  -(s \cdot k)(q \cdot Q)) 
\end{eqnarray*} 
\vspace*{-0.4in}
\begin{equation}
\hspace*{-2.28in}
  +4\Im (\Lambda f_{S}f_{V}^{*})\epsilon_{\beta\alpha\mu\rho}
      s^{\beta}q^{\alpha}Q^{\mu}k^{\rho}, \label{eq:A10}
\end{equation}

\begin{eqnarray*}
|\bar{M}^{2}| & \sim & 2|f_{V}|^2(2(q \cdot Q)(k \cdot Q)-(q \cdot k)Q^2)
                       + 2|\Lambda|^2|f_{S}|^2(q \cdot k) 
~~~~~~~~~~~~~~~~~~~~~~~~~~~~~~\\
              && +4\Re(\Lambda f_{S}f_{V}^{*})m(Q \cdot k)\\
              &&
  -2|f_{V}|^2m(2(s \cdot Q)(k \cdot Q) - (s \cdot k)Q^2)
  +2|\Lambda|^2|f_{S}|^2m(s \cdot k)\\
              && 
  -4\Re(\Lambda f_{S}f_{V}^{*})((s \cdot Q)(q \cdot k) 
  -(s \cdot k)(q \cdot Q))
\end{eqnarray*}
\vspace*{-0.4in}
\begin{equation}
\hspace*{-2.3in}
  +4\Im (\Lambda f_{S}f_{V}^{*})\epsilon_{\beta\alpha\mu\rho}
       s^{\beta}q^{\alpha}Q^{\mu}k^{\rho}. \label{eq:A11}
\end{equation}
Combining these two equations, dropping the overall factor of two and 
using $\Lambda^{+} \equiv \Lambda$ for $\tau^{-}$ and 
$\Lambda^{+} \equiv \Lambda^{*}$ for $\tau^{+}$ gives us the following
expression for the squared matrix element of 
$\tau^\pm \to \pi^\pm \pi^0 \nu_{\tau}$:
\begin{eqnarray*}
|M^{2}| &\sim & |f_{V}|^2(2(q \cdot Q)(k \cdot Q)-(q \cdot k)Q^2) 
+ |\Lambda|^2|f_{S}|^2(q \cdot k) ~~~~~~~~~~~~~~~~~~~~~~~~~~~~~~~~\\
        &&+ 2\Re(\Lambda^+ f_{S}f_{V}^{*})m(Q \cdot k)\\
        &&   \pm |f_{V}|^2m(2(s \cdot Q)(k \cdot Q)-(s \cdot k)Q^2)
             \mp |\Lambda|^2|f_{S}|^2m(s \cdot k) \\ 
        &&   \pm 2\Re(\Lambda^{+}f_{S}f_{V}^{*})((s \cdot Q)(q \cdot k) 
              +(s \cdot k)(q \cdot Q))
\end{eqnarray*}
\vspace*{-0.4in}
\begin{equation} 
\hspace*{-2.16in}
        + 2\Im (\Lambda^{+}f_{S}f_{V}^{*})\epsilon_{\beta\alpha\mu\rho}
          s^{\beta}q^{\alpha}Q^{\mu}k^{\rho}.
\label{eq:rate_not_simplified}
\end{equation}
We can separate the CP-even and odd part in the squared matrix element
by using the following equivalents:
\begin{equation}
 \Re(\Lambda^+ f_S f_V^*) = 
               \Re(\Lambda)|f_v||f_s|~\cos\delta \mp 
                             \Im (\Lambda)|f_v||f_s|\;\sin\delta,
\end{equation}
\begin{equation}
 \Im (\Lambda^+ f_S f_V^*) = 
               \Re(\Lambda)|f_v||f_s|\;\sin\delta \pm 
                         \Im (\Lambda)|f_v||f_s|\;\cos\delta.
\end{equation}
Now we can rewrite Eq.~\ref{eq:rate_not_simplified} as:
\begin{eqnarray*}
|{\cal M}|^2_{\tau^\pm \to \pi^\pm\pi^0\nu} & \sim &
       |f_v|^2[2(qQ)(kQ)-(kq)Q^2] + |\Lambda|^2|f_s|^2(qk) \\
&&     2\Re(\Lambda) |f_v||f_s|\;\cos\delta ~M_{\tau} (Qk) -
    2\Im (\Lambda^+) |f_v||f_s|\;\sin\delta ~M_{\tau} (Qk) \\
&&  + s^\mu \{\pm |f_v|^2 M_{\tau}(2Q_{\mu}(kQ)-k_{\mu}Q^2)
    \mp |\Lambda|^2 |f_s|^2 M_\tau k_\mu \\
&&  \pm 2\Re(\Lambda)|f_v||f_s|\;\cos\delta ~(Q_\mu(kq)-k_\mu (qQ)) \\ 
&&  \mp 2\Im (\Lambda^+)|f_v||f_s|\;\sin\delta~(Q_\mu(kq)-k_\mu (qQ))\\
&&  + 2\Re(\Lambda)|f_v||f_s|\;\sin\delta~
        e_{\mu\alpha\beta\gamma}q^\alpha Q^\beta k^\gamma
\end{eqnarray*}
\vspace*{-0.4in}
\begin{equation}   
           + 2\Im (\Lambda^+)|f_v||f_s|\;\cos\delta~
        e_{\mu\alpha\beta\gamma}q^\alpha Q^\beta k^\gamma \}.
\label{eq:rate}
\end{equation}
\section{\boldmath $\tau$-pair production and spin correlations}

The matrix element of $e^+e^-\to\tau^+\tau^-$ can be written as:
\begin{equation}
M \sim [\bar{u(\tau^-)}\gamma^\mu v(\tau^+)][\bar{v}(e^+)\gamma_{\mu}u(e^-)].
\end{equation}
We denote the four-momenta of $\tau^\pm$ as $q_1$ and $q_2$, their spins
as $s_1$ and $s_2$, respectively, the mass of tau lepton as $m$ and the 
four-momenta of positron as $p_1$ and 
electron as $p_2$. We then write the squared matrix element as:
\begin{equation}
|M|^2 \sim Tr(\slashchar{p}_1\gamma_\mu \slashchar{p}_2\gamma_\nu)
           Tr((1-\gamma_5\slashchar{s}_2)(\slashchar{q}_2+m)\gamma^\mu
              (1-\gamma_5\slashchar{s}_1)(\slashchar{q}_1-m)\gamma^\nu).
\end{equation}
Calculating the traces we get the following expression:
\begin{eqnarray*}
 |M|^2 & \sim & (p_1\cdot q_1)(p_2\cdot q_2) + (p_1 \cdot q_2)(p_2\cdot q_1)
             + m^2(p_1 \cdot p_2) \\
&& -(s_1\cdot s_2)(p_1\cdot q_2)(p_2\cdot q_1) 
   -(s_1\cdot s_2)(p_1\cdot q_1)(p_2\cdot q_2)
   +(s_1\cdot s_2)(p_1\cdot p_2)(q_1\cdot q_2) \\
&& -m^2(s_1 \cdot p_2)(s_2 \cdot p1) -m^2 (s_1 \cdot p_1)(s_2 \cdot p_2)\\
&&   -(p_1 \cdot p_2)(s_1 \cdot q_2)(s_2 \cdot q_1)
     -\underline{(p_1 \cdot p_2)(s_1 \cdot q_1)(s_2 \cdot q_2)} \\
&&   +(p_1 \cdot q_1)(q_2 \cdot s_1)(p_2 \cdot s_2)
     +(p_1 \cdot q_2)(p_2 \cdot s_1)(q_1 \cdot s_2)\\
&&   +(p_2 \cdot q_1)(q_2 \cdot s_1)(p_1 \cdot s_2)
     +(p_2 \cdot q_2)(p_1 \cdot s_1)(q_1 \cdot s_2)
\end{eqnarray*}
\vspace*{-0.5in}
\begin{equation}
\hspace*{-0.88in}
     -(q_1 \cdot q_2)(p_2 \cdot s_1)(p_1 \cdot s_2)
     -(q_1 \cdot q_2)(p_1 \cdot s_1)(p_2 \cdot s_2). 
\end{equation}
The gauge condition $(q_i\cdot s_i) = 0$ leads to a vanishing underlined term.
Re-writing the squared matrix element in terms of
spin-averaged production $P$ and spin-dependent part 
$s_{1\mu}s_{2\nu}C^{\mu\nu}$ we obtain the explicit form of
the spin-correlation matrix:
\begin{eqnarray*}
|M|^2 & \sim & P + s_{1\mu}s_{2\nu}C^{\mu\nu} = \\
&& (p_1\cdot q_1)(p_2\cdot q_2) + (p_1 \cdot q_2)(p_2\cdot q_1)
             + m^2(p_1 \cdot p_2) \\
&& + s_{1\mu}s_{2\nu}\{-g^{\mu\nu}
     [(p_1\cdot q_2)(p_2\cdot q_1)+(p_1\cdot q_1)(p_2\cdot q_2)-
      (p_1\cdot p_2)(q_1\cdot q_2)] \\
&& \mbox{~~~~~~~~~~~}   - m^2 p_2^\mu p_1^\nu -m^2 p_1^\mu p_2^\nu
        -(p_1\cdot p_2)q_2^\mu q_1^\nu + (p_1\cdot q_1)q_2^\mu p_2^\nu \\
&& \mbox{~~~~~~~~~~~}+ (p_1\cdot q_2)p_2^\mu q_1^\nu
                     + (p_2\cdot q_1)q_2^\mu p_1^\nu
                     + (p_2\cdot q_2)p_1^\mu q_1^\nu 
\end{eqnarray*}
\vspace*{-0.5in}
\begin{equation}
\hspace*{-0.4in}
- (q_1\cdot q_2)p_2^\mu p_1^\nu
- (q_1\cdot q_2)p_1^\mu p_2^\nu \},
\end{equation}
or
\begin{eqnarray*}
C^{\mu\nu} & = & -g^{\mu\nu}
     [(p_1\cdot q_2)(p_2\cdot q_1)+(p_1\cdot q_1)(p_2\cdot q_2)-
      (p_1\cdot p_2)(q_1\cdot q_2)] \\
&&    - m^2 p_2^\mu p_1^\nu -m^2 p_1^\mu p_2^\nu
      -(p_1\cdot p_2)q_2^\mu q_1^\nu + (p_1\cdot q_1)q_2^\mu p_2^\nu \\
&&    +(p_1\cdot q_2)p_2^\mu q_1^\nu
      +(p_2\cdot q_1)q_2^\mu p_1^\nu
      +(p_2\cdot q_2)p_1^\mu q_1^\nu
\end{eqnarray*}
\vspace*{-0.5in}
\begin{equation}
\hspace*{-1.18in}
    -(q_1\cdot q_2)p_2^\mu p_1^\nu
      - (q_1\cdot q_2)p_1^\mu p_2^\nu. 
\end{equation}
The explicit form of the spin-correlation matrix is equal to:
\begin{equation}
  \tilde{C}_{\mu\nu} = 
    [g_{\mu\alpha} - \frac{1}{m_{\tau}^2}q_{\mu}q_{\alpha}]
                         \: C^{\alpha\beta} \:
    [g_{\beta\nu} - \frac{1}{m_{\tau}^2}\bar{q}_{\beta}\bar{q}_{\nu}]. 
\label{eq:Ctilde}
\end{equation}
%
%
%
%
%
%
%
%
%
%
%
%
\section{\boldmath 
         CP-even and odd parts of tau pair correlated decay rate into 
         $(\pi^-\pi^0\nu_{\tau})-(\pi^+\pi^0\bar{\nu}_{\tau})$}

The total probability density for the tau pair decays is given by 
Eq.~\ref{eq:Ptotal}:
\begin{equation}
P_{\tau^+\tau^-\to\pi^-\pi^0\nu~\pi^+\pi^0\bar{\nu}} = 
G\times \sigma \times \bar{G} + \omega_\mu \tilde{C}^{\mu\nu} \bar{\omega_\nu},
\end{equation}
where $G$, $\bar{G}$ are the spin-averaged matrix elements squared 
for $\tau^-$, $\tau^+$ decay (Eq.~\ref{eq:Mtau_decay}), 
$\sigma = (p\bar{q})^2 + (\bar{p}\bar{q})^2+m_{\tau}^2(\bar{p}p)$ is the
spin-averaged cross section of $e^+e^-\to\tau^+\tau^-$, $\omega^{\mu}$ and
$\bar{\omega}^{\nu}$ are the polarimeter vectors for $\tau^-$ and $\tau^+$
(Eq.~\ref{eq:Mtau_decay3}), and $\tilde{C}_{\mu\nu}$ is the 
spin-correlation matrix (Eq.~\ref{eq:Ctilde}).
Both the spin-averaged matrix element and polarimeter vector of each tau decay
contain CP-odd and CP-even terms. Those terms are contracted with the CP-even
tau pair production cross section and spin-correlation matrix. Therefore,
the total CP-odd part of Eq.~\ref{eq:Ptotal} is a product of the
CP-even terms of one tau decay contracted with the CP-odd terms
of the other tau decay, $i.e.$, it is a linear function of the CP-odd terms. 
Similarly, the CP-even part of Eq.~\ref{eq:Ptotal} is equal to the CP-even 
terms of one tau decay contracted with the CP-even terms of the other 
tau plus the the CP-odd terms of one tau decay contracted with the CP-odd ones 
from the other tau decay. This is a sum of the non-CP-violating terms of 
tau decays and the contraction of the CP-violating terms for both decays 
simultaneously, where each term remains CP-even.
The matrix element for the tau decay is given by Eqs.~\ref{eq:Mtau_decay},
~\ref{eq:Mtau_decay2}, and \ref{eq:Mtau_decay3}, where the terms underlined 
are CP-odd. Thus, the CP-even term of the total cross section is equal to:\\
$     P_{even} = $
$$            \{ |f_v|^2[2(qQ)(kQ)-(kq)Q^2] + |\Lambda|^2|f_s|^2(qk) + 
                2\Re(\Lambda)|f_v||f_s|\;\cos\delta ~M_{\tau} (Qk)\}$$
$$		\times \sigma \times $$
$$                \{ |\bar{f}_v|^2[2(\bar{q}\bar{Q})
                 (\bar{k}\bar{Q})-(\bar{k}\bar{q})\bar{Q}^2] 
                 + |\Lambda|^2|\bar{f}_s|^2(\bar{q}\bar{k}) +
                 2\Re(\Lambda) |\bar{f}_v||\bar{f}_s|\;\cos\bar{\delta}~M_{\tau}(\bar{Q}\bar{k}) \}$$
$	+$
%
%
$$           \underline{(-2\Im (\Lambda)|f_v||f_s|\;\sin\delta ~M_{\tau} (Qk))}
		\times \sigma \times
             (\underline{-2\Im (\Lambda^*)|\bar{f}_v||\bar{f}_s| 
                 \;\sin\bar{\delta} ~M_{\tau} (\bar{Q}\bar{k})}) $$
$	+$
%
%
$$    \{ |f_v|^2 M_{\tau}(2Q_{\mu}(kQ)-k_{\mu}Q^2)
        -|\Lambda|^2 |f_s|^2 M_\tau k_\mu $$
$$      + 2\Re(\Lambda)|f_v||f_s|\;\cos\delta~(Q_\mu(kq)-k_\mu (qQ)) 
        + 2\Re(\Lambda)|f_v||f_s|\;\sin\delta~
         e_{\mu\alpha\beta\gamma}q^\alpha Q^\beta k^\gamma \}$$
$$	\cdot \tilde{C}^{\mu\nu} \cdot$$
$$    \{ - |\bar{f}_v|^2 M_{\tau}
           (2\bar{Q}_{\mu}(\bar{k}\bar{Q})-\bar{k}_{\mu}\bar{Q}^2)
         + |\Lambda|^2 |\bar{f}_s|^2 M_\tau \bar{k}_\mu $$
$$       - 2\Re(\Lambda)|\bar{f}_v||\bar{f}_s|\;\cos\bar{\delta}~
             (\bar{Q}_\mu(\bar{k}\bar{q})-\bar{k}_\mu (\bar{q}\bar{Q})) 
         + 2\Re(\Lambda)|\bar{f}_v||\bar{f}_s|\;\sin\bar{\delta}~
         e_{\nu\alpha\beta\gamma}\bar{q}^\alpha\bar{Q}^\beta\bar{k}^\gamma\}$$
$	+$
%
%
$$  \{ \underline{-2\Im (\Lambda)|f_v||f_s|\;\sin\delta~(Q_\mu(kq)-k_\mu (qQ))}
            +\underline{2\Im (\Lambda)|f_v||f_s|\;\cos\delta~
        e_{\mu\alpha\beta\gamma}q^\alpha Q^\beta k^\gamma } \}$$
$$	\cdot \tilde{C}^{\mu\nu} \cdot$$
\begin{equation}
\label{eq:Peven}
  \{ \underline{2\Im (\Lambda^*)|\bar{f}_v||\bar{f}_s|\;\sin\bar{\delta}~
         (\bar{Q}_\mu(\bar{k}\bar{q})-\bar{k}_\mu (\bar{q}\bar{Q}))}   
           + \underline{2\Im (\Lambda^*)|\bar{f}_v||\bar{f}_s|\;\cos\bar{\delta}~
        e_{\nu\alpha\beta\gamma}\bar{q}^\alpha\bar{Q}^\beta\bar{k}^\gamma} \}.
\end{equation}

\noindent Similarly, the expression for CP-odd part of the tau pair 
production decaying into $\pi\pi^0\nu_{\tau}$ state is:\\
$ P_{odd} = $
$$            \{ |f_v|^2[2(qQ)(kQ)-(kq)Q^2] + |\Lambda|^2|f_s|^2(qk) + 
                2\Re(\Lambda)|f_v||f_s|\;\cos\delta ~M_{\tau} (Qk)\}$$
$$		\times \sigma \times
             (\underline{(-2\Im (\Lambda^*)|\bar{f}_v||\bar{f}_s|
               \;\sin\bar{\delta} ~M_{\tau} (\bar{Q}\bar{k})})) $$
$	+$
$$             \underline{(-2\Im (\Lambda)|f_v||f_s|\;\sin\delta ~M_{\tau}(Qk))}
		\times \sigma \times $$
$$                \{ |\bar{f}_v|^2[2(\bar{q}\bar{Q})
                 (\bar{k}\bar{Q})-(\bar{k}\bar{q})\bar{Q}^2] 
                 + |\Lambda|^2|\bar{f}_s|^2(\bar{q}\bar{k}) + 
                 2\Re(\Lambda) |\bar{f}_v||\bar{f}_s|\;\cos\bar{\delta}~M_{\tau}(\bar{Q}\bar{k}) \}$$
$	+$
$$    \{ |f_v|^2 M_{\tau}(2Q_{\mu}(kQ)-k_{\mu}Q^2)
        -|\Lambda|^2 |f_s|^2 M_\tau k_\mu $$
$$      + 2\Re(\Lambda)|f_v||f_s|\;\cos\delta~(Q_\mu(kq)-k_\mu (qQ)) 
        + 2\Re(\Lambda)|f_v||f_s|\;\sin\delta~
         e_{\mu\alpha\beta\gamma}q^\alpha Q^\beta k^\gamma \}$$
$$	\cdot \tilde{C}^{\mu\nu} \cdot$$
$$  \{ \underline{2\Im (\Lambda^*)|\bar{f}_v||\bar{f}_s|\;\sin\bar{\delta}~
         (\bar{Q}_\mu(\bar{k}\bar{q})-\bar{k}_\mu (\bar{q}\bar{Q}))}   
           + \underline{2\Im (\Lambda^*)|\bar{f}_v||\bar{f}_s|\;\cos\bar{\delta}~
        e_{\nu\alpha\beta\gamma}\bar{q}^\alpha\bar{Q}^\beta\bar{k}^\gamma} \}$$
$	+$
%
%
$$  \{ \underline{-2\Im (\Lambda)|f_v||f_s|\;\sin\delta~(Q_\mu(kq)-k_\mu (qQ))}
            +\underline{2\Im (\Lambda^*)|f_v||f_s|\;\cos\delta~
        e_{\mu\alpha\beta\gamma}q^\alpha Q^\beta k^\gamma } \}$$
$$	\cdot \tilde{C}^{\mu\nu} \cdot$$
$$    \{ - |\bar{f}_v|^2 M_{\tau}
           (2\bar{Q}_{\mu}(\bar{k}\bar{Q})-\bar{k}_{\mu}\bar{Q}^2)
         + |\Lambda|^2 |\bar{f}_s|^2 M_\tau \bar{k}_\mu $$
\begin{equation}
\label{eq:P_odd}
       - 2\Re(\Lambda)|\bar{f}_v||\bar{f}_s|\;\cos\bar{\delta}~
             (\bar{Q}_\mu(\bar{k}\bar{q})-\bar{k}_\mu (\bar{q}\bar{Q})) 
         + 2\Re(\Lambda)|\bar{f}_v||\bar{f}_s|\;\sin\bar{\delta}~
         e_{\nu\alpha\beta\gamma}\bar{q}^\alpha\bar{Q}^\beta\bar{k}^\gamma\}.
\end{equation}
%
%
%
%
%
%
\section{The optimal observable}
There is a freedom of choice of the CP sensitive observable 
$\xi$ used in our search.\footnote{We assume that the observable is CP-odd.} 
To maximize the sensitivity of the measurement we need to select 
$\xi$ with a minimal relative statistical error. For $N$ events, 
the statistical error on the average value  $<\xi>$ is given by:
\begin{equation}
 \Delta \xi = \sqrt{\frac{<\xi^2> - <\xi>^2}{N}}. 
\end{equation}
If the contribution of the CP-odd term is small ($i.e.,~\Im(\Lambda)$ is small),  
the $<\xi>^2$ term is proportional to $|\Im(\Lambda)|^2$ 
(see Eq.~\ref{eq:xi_ave}) and can be neglected. Therefore:
\begin{equation}
 \Delta \xi \sim \sqrt{\frac{<\xi^2>}{N}} = 
              \sqrt{\frac{\int \xi^2  P dLips}{N~\Gamma}},
 \label{eq:error}
\end{equation}  
where $P$ is a probability density of the process and $\Gamma$ 
is a normalization factor, equal to:
\begin{equation}
 \Gamma = \int P ~dLips. 
\end{equation} 
The sensitivity $S$ of this method to the presence of a non-zero
value of $\Im(\Lambda)$ is measured by the number of standard deviations
of $<\xi>$ with respect to 
zero:\footnote{The average of the observable $\xi$ is equal to zero if there 
is no CP violation.} 
\begin{equation}
\label{eq:sens}
 S = \frac{<\xi>}{\Delta \xi} = 
    \frac{\int \xi ~P ~dLips}{\sqrt{\frac{<\xi^2> - <\xi>^2}{N}}}.
\end{equation}
We can simplify this expression by separating the CP-odd and CP-even parts 
of the probability density $P$ (see Eq.~\ref{eq:Ptotal2}):
\begin{equation}
 P = P_{even} +  P_{odd}.
\end{equation}
The sensitivity becomes:
\begin{equation}
  S =   \sqrt{N~\Gamma R}, 
  {\rm~where~} R = \frac{(\int \xi ~ P_{odd} ~dLips)^2}
                   {\int \xi^2 (P_{even}+  P_{odd})dLips}.
 \label{eq:sign}
\end{equation}  
We can further simplify $R$ by using the fact that $\xi^2 P_{odd}$ is a CP-odd 
function:
\begin{equation}
\int \xi^2 ~ P_{odd} ~dLips = 0,
\end{equation} 
and
\begin{equation}
 R = \frac{(\int \xi  P_{odd} ~dLips)^2}{\int \xi^2 P_{even} ~dLips}.
\label{eq:R}
\end{equation}
The only $\xi$ dependent factor in the expression for the sensitivity is R. 
In order to find an observable which is most sensitive to the CP-odd term,
we need to find a function $\xi$, for which the value of $R$ is maximal.
We use the functional differentiation method. R is in extremum, when its 
first derivative with respect to $\xi$ is equal to zero:
\begin{equation}
\frac{\delta R}{\delta \xi} = 
\frac{2P_{odd}\int\xi~~P_{odd}~dLips\int\xi^2~P_{even}~dLips - 
      2\xi~P_{even}(\int\xi~~P_{odd}~dLips)^2}
                  {(\int\xi^2~P_{even}~dLips)^2} 
      = 0,
\label{eq:extremum1}
\end{equation}
or:
\begin{equation}
P_{odd}\int\xi~~P_{odd}~dLips\int\xi^2~P_{even}~dLips - 
      \xi~P_{even}(\int\xi~~P_{odd}~dLips)^2 = 0.
\label{eq:extremum2}
\end{equation}
If $\int\xi~~P_{odd}~dLips=0$, then the condition above is satisfied,
but the sensitivity is equal to zero (see Eq.~\ref{eq:sign}), thus
such case is not a proper solution. 
If $\int\xi~~P_{odd}~dLips\ne0$, then we can simplify 
Eq.~\ref{eq:extremum2} to obtain:
\begin{equation}
P_{odd}\int\xi^2~P_{even}~dLips = \xi~P_{even}\int\xi~~P_{odd}~dLips.
\label{eq:extremum3}
\end{equation}
One of the possible condition for the extremum of $R$ is:
\begin{equation}
\xi_{ext} = A\frac{ P_{odd}}{P_{even}},
\label{eq:first_solution}
\end{equation}
where $A$ is an arbitrary constant. The constant $A$ must be real for 
the variable $\xi$ to be CP-odd. We can decompose any CP-odd observable 
$\xi$ into $\xi_{ext}$ and the reminder $\hat{\xi} \equiv \xi - \xi_{ext}$:
\begin{equation}
 \xi = A\frac{P_{odd}}{P_{even}} + \hat{\xi}. 
\end{equation}
Substituting the value of $\xi$ into Eq.~\ref{eq:R} gives us the 
following result:
\begin{equation}
R = \frac{(\int A\frac{P_{odd}^2}
                {P_{even}}~dLips+\int\hat{\xi}P_{odd} ~dLips)^2}
    {\int A^2\frac{P_{odd}^2}{\Omega}~dLips+
 2A\int P_{odd} \hat{\xi} ~dLips+\int \hat{\xi}^2 P_{even} ~dLips}.
\label{eq:R_simplified}
\end{equation}
We can simplify this equation using that fact that value of $R$ does 
not depend  on a constant $A$ (see Eq.~\ref{eq:first_solution}). 
We define $A$ such that:
\begin{equation}
     \int P_{odd} \hat{\xi} ~dLips = 0.
     \label{eq:ksi_hat}
\end{equation}
Then, the numerator and one of the terms of the denominator of 
Eq.~\ref{eq:R_simplified} vanish. To determine 
the value of $A$, we multiply left and right sides of 
Eq.~\ref{eq:first_solution} by $P_{odd}$ and integrate over the 
phase space:
\begin{equation}
\int \hat{\xi} P_{odd} ~dLips = A \int \frac{P_{odd}^2}{P_{even}} ~dLips +
                          \int \hat{\xi} P_{odd} ~dLips,  
\end{equation}
\begin{equation}
 \int \hat{\xi} P_{odd} ~dLips = \int \xi P_{odd} ~dLips -  
               A \int \frac{P_{odd}^2}{P_{even}} ~dLips =0 , {\rm~if:}
\end{equation}
\begin{equation}
A = \frac{\int \xi P_{odd} ~dLips}{\int \frac{P_{odd}^2}{P_{even}} ~dLips}.
\label{eq:A}
\end{equation}
Choosing $A$ in this way, we can simplify $R$, using the property of 
Eq.~\ref{eq:ksi_hat}:
\begin{equation}
R = \frac{( \int \frac{P_{odd}^2}{P_{even}}~dLips)^2}
      {\int\frac{P_{odd}^2}{P_{even}}~dLips + \int\hat{\xi}^2 P_{odd} ~dLips}.
\label{eq:R2}
\end{equation}
The numerator $(\int\frac{P_{odd}^2}{P_{even}}~dLips)^2$ 
and the first term of the denominator $\int\frac{P_{odd}^2}{P_{even}}~dLips$ 
are positive and do not depend on $\hat{\xi}$.
$R$ has a maximal value when the last term in the denominator is 
equal to zero. This is possible only when $\hat{\xi}$ is equal to zero. 
Therefore, the value of $R$ is at maximum for $\xi = A\frac{P_{odd}}{P_{even}}$, 
and $\xi$ is most sensitive to the CP-odd part of the decay rate. 
Since the sensitivity does not depend on a value of the constant $A$,
we choose $A=1$ corresponding to:
\begin{equation}
\xi = \frac{P_{odd}}{P_{even}}.
\label{eq:opt}
\end{equation}
%
%
%


\begin{thebibliography}{99}
\bibitem{matter}{A.D.~Sakharov, ZhETF Pis. Red. {\bf 5}, 32 (1967); 
JETP Lett. {\bf 5}, 24 (1967).}
\bibitem{cpk}{J.H.~Christenson, J.W.~Cronin, V.L.~Fitch and R.~Turlay, 
Phys. Rev. Lett. {\bf 13}, 138 (1964).}
\bibitem{direct}{NA48 Collaboration, V.~Fanti $et~al.$, Phys. Rev. B {\bf 465}, 335 (1999);
 KTeV Collaboration, A.~Alavi-Harati $et~al.$, Phys. Rev. Lett {\bf 83}, 22 (1999).}
\bibitem{kpimu1}{KEK-E246 Collaboration, M.~Abe $et~al.$, Phys. Rev. Lett. {\bf 83}, 4253 (1999).}
\bibitem{kpimu2}{S.R.~Blatt $et~al.$, Phys. Rev. D {\bf 27}, 1056 (1983).}
\bibitem{bigi}{I.I.~Bigi and A.I.~Sanda, {\it CP Violation}, 
(Cambridge University Press 1999).}
\bibitem{BABAR}{B.~Aubert $et~al$. Phys. Rev. Lett. {\bf 86}, 2515 (2001).}
\bibitem{BELLE}{A.~Abashian $et~al$. Phys. Rev. Lett. {\bf 86}, 2509 (2001).}
\bibitem{kamiokande}{Kamiokande Collaboration, T. Kajita $et~al.$, Nucl. Phys. Proc. Suppl. {\bf 77},
123 (1999).}
\bibitem{wudka}{M.B.~Einhorn, J.~Wudka, hep-ph/0007285.}
\bibitem{lfv1}{K.~Matsuda, N.~Takeda, T.~Fukuyama, H.~Nishiura, Phys. Rev. D {\bf 62} 093001 (2000).}
\bibitem{lfv2}{R.~Kitano and Y.~Okada, hep-ph/0012040.}
\bibitem{mhdm}{Y.~Grossman, Nucl. Phys. {\bf B426}, 355 (1994).}
\bibitem{mhdm2}{Y.~Grossman, Y.~Nir, R.~Rattazzi,  SLAC-PUB-7379, hep-ph 9701231.}
\bibitem{mhdm3}{C.H.~Albright, J.~Smith, S.H.H.~Tye, Phys. Rev. D {\bf 21}, 711 (1980).}
\bibitem{tsai}{Y.S.~Tsai, Nucl. Phys. Proc. Suppl. {\bf 55C}, 293 (1997).}
\bibitem{kuhn}{J.H.~Kuhn, E.~Mirkes, Phys. Lett. B {\bf 398}, 407 (1997).}
\bibitem{mu_decays_work}{H.~Burkhardt $et~al.$, Phys. Lett. B {\bf 160}, 343, (1985).}
\bibitem{mu_decays}{D.M.~Kaplan, Phys. Rev. D {\bf 57}, 3827 (1998), and
references therein.} 

\bibitem{colin}{CLEO Collaboration, S.~Anderson $et~al.$, Phys. Rev. Lett. {\bf 81}, 3823, (1998).}
\bibitem{belle_result}{BELLE Collaboration, {\it CP/T tests with $\tau$ leptons at Belle}, to be published in the proceedings of 30th International Conference on High-Energy Physics (ICHEP 2000), Osaka, Japan, 27 Jul - 2 Aug 2000.} 
\bibitem{3hdm1}{S.~Weinberg, Phys. Rev. Lett. {\bf 37}, 657 (1976).}
\bibitem{tsai2}{Y.S.~Tsai, Phys. Rev. D {\bf 51}, 3172, (1995).}
\bibitem{alan}{We thank A.~Weinstein and F.~Olness for help in this 
calculation.}
\bibitem{rhoprime}{CLEO Collaboration, S.~Anderson $et al.$, Phys. Rev. D {\bf 61}, 112002, (2000).}
\bibitem{optim1}{D.~Atwood and A.~Soni, Phys. Rev. D {\bf 45}, 2405 (1992).}
\bibitem{optim2}{J.~Gunion, B.~Grzadkowski, X.~He, Phys. Rev. Lett. {\bf 77}, 5172 (1996).}
\bibitem{KORALB}{KORALB (v.~2.2)/TAUOLA (v.~2.4): S.~Jadach, J.H.~Kuhn and
Z.~Was, Comput. Phys. Commun. {\bf 36}, 191 (1985) and {\it ibid} {\bf 64},
275 (1991), {\it ibid} {\bf 76}, 361 (1993). Implementation of the scalar
in $\tau\to \pi\pi^0\nu_{\tau}$ decay is done by A.~Weinstein 
and M.~Schmidtler and can be obtained directly from them (ajw@caltech.edu).}
\bibitem{cone}{CLEO Collaboration, R.Balest $et~al.$, Phys. Rev. D {\bf 47}, 3671 (1993).}
\bibitem{CLEOII}{CLEO Collaboration, Y.~Kubota $et~al.$, Nucl. Instr. and Meth. {\bf A320}, 66 (1992).}
\bibitem{CLEOII5}{CLEO Collaboration, T.S.~Hill, Nucl. Instr. and Meth. {\bf A418}, 32 (1998).}
\bibitem{GEANT}{R.~Brun $et~al.$, GEANT 3.15, CERN DD/EE/84-1.}
\bibitem{gamgam}{V.M.~Budnev $et~al.$, Phys. Rep. {\bf C15}, 181 (1975).}
\bibitem{rho-rho}{CLEO Collaboration, M.~Artuso $et~al.$, Phys. Rev. Lett. {\bf 73}, 3762 (1994).}
\bibitem{PDG_leptonic}{Particle Data Group, D.E.~Groom $et~al.$, 
Eur. Phys. J. {\bf C15}, 320 (2000).}

\end{thebibliography}
\end{document}